\documentclass{sig-alternate-05-2015}
\usepackage{booktabs} 
\usepackage{subcaption}
\usepackage{hyperref}
\begin{document}
\title{Focused Crawl of Web Archives to Build Event Collections}
\numberofauthors{3}
\author{
%
\alignauthor
Martin Klein \\
        \affaddr{Research Library}\\
        \affaddr{Los Alamos National Laboratory}\\
        \affaddr{Los Alamos, NM, USA}\\
        \affaddr{\url{http://orcid.org/0000-0003-0130-2097}}\\
        \email{mklein@lanl.gov}
\and
\alignauthor
Lyudmila Balakireva\\
        \affaddr{Research Library}\\
        \affaddr{Los Alamos National Laboratory}\\
        \affaddr{Los Alamos, NM, USA}\\
        \affaddr{\url{http://orcid.org/0000-0002-3919-3634}}\\
        \email{ludab@lanl.gov}
\alignauthor
Herbert Van de Sompel\\
        \affaddr{Research Library}\\
        \affaddr{Los Alamos National Laboratory}\\
        \affaddr{Los Alamos, NM, USA}\\
        \affaddr{\url{http://orcid.org/0000-0002-0715-6126}}\\
        \email{herbertv@lanl.gov}
}
\maketitle
\begin{abstract}
Event collections are frequently built by crawling the live web on the basis of seed URIs nominated by human 
experts. Focused web crawling is a technique where the crawler is guided by reference content pertaining 
to the event.
Given the dynamic nature of the web and the pace with which topics evolve, the timing of the crawl
is a concern for both approaches. 
We investigate the feasibility of performing focused crawls on the archived web. By utilizing the Memento 
infrastructure, we obtain resources from $22$ web archives that contribute to building event collections.
We create collections on four events and compare the relevance of their resources to collections built from
crawling the live web as well as from a manually curated collection.
Our results show that focused crawling on the archived web can be done and indeed results in highly relevant
collections, especially for events that happened further in the past.
\end{abstract}
\section{Introduction}
The pace at which real-world events happen paired with the level of event coverage on the web has by
far outgrown the human capacity for information consumption. Therefore, archivists and librarians are
interested in building special event-centric web collections that humans can consult post-factum.
Web crawling on the basis of seed URIs is a common approach to collect such event-specific web resources.
For example, the Archive-It service\footnote{\url{https://archive-it.org/}} is frequently used to crawl the web
to build archival collections on the basis of seeds URIs\footnote{\url{https://twitter.com/archiveitorg/status/960564121577181184}}\footnote{\url{https://twitter.com/internetarchive/status/806228431474028544}}\footnote{\url{https://twitter.com/internetarchive/status/797263535994613761}}
that were manually collected by librarians, archivists, and volunteers. This approach has drawbacks since the notion 
of relevance is solely based on the nomination of seed URIs by humans. Focused web crawling guided by a set of 
reference documents that are exemplary of the web resources that should be collected is an approach that is commonly 
used to build special-purpose collections. It entails an algorithmic assessment of the relevance of the content of
a crawled resource rather than a manual selection of URIs to crawl. For both web crawling and focused web crawling, 
the time between the occurrence of the event and the start of the crawling process is a concern since stories 
quickly disappear from the top search engine result pages \cite{nwala:scaping_serps}, links rot, and content 
drifts \cite{klein:one_in_five,jones:content_drift}.
Web archives around the world routinely collect snapshots of web pages (which we
refer to as Mementos) and hence potentially are repositories from which event-specific collections
could be gathered some time after the event. However, the various web archives have different scopes e.g.,
national vs. international resources, cover different time spans, and vary in size of their
index\footnote{\url{https://twitter.com/brewster_kahle/status/954889200083509248}}. This makes collection
building on the basis of distributed web archives difficult when compared to doing so on the live web.
Moreover, to the best of our knowledge, focused crawling across web archives has never been attempted.
Inspired by previous work by Gossen et al. \cite{gossen:extracting}, in this paper,
we present a framework to build event-specific collections by focused crawling of web archives.
We utilize the Memento protocol \cite{memento} and the associated cross-web-archive
infrastructure \cite{Bornand:MementoML} to crawl Mementos in $22$ web archives. We build collections 
by evaluating the content-wise and temporal relevance of crawled resources and we compare the resulting 
collections with collections created on the basis of live web crawls and a manually curated Archive-It crawl.
As such, we take the previous work to the next level and ask the following questions:
\begin{itemize}
\item Can we create event collections by focused crawling web archives?
\item How do event collections created from the archived web compare to those created from the live web?
\item How does the amount of time passed since the event affect the collections built from the live and 
the archived web?
\item How do event collections built from the archived web compare to manually curated collections?
\end{itemize}
We consider the main contribution of our work to be the exploration of the feasibility of
performing focused crawls on the archived web. To the best of our knowledge,
we are the first to do so.
\begin{figure}[t!]
\includegraphics[scale=0.45]{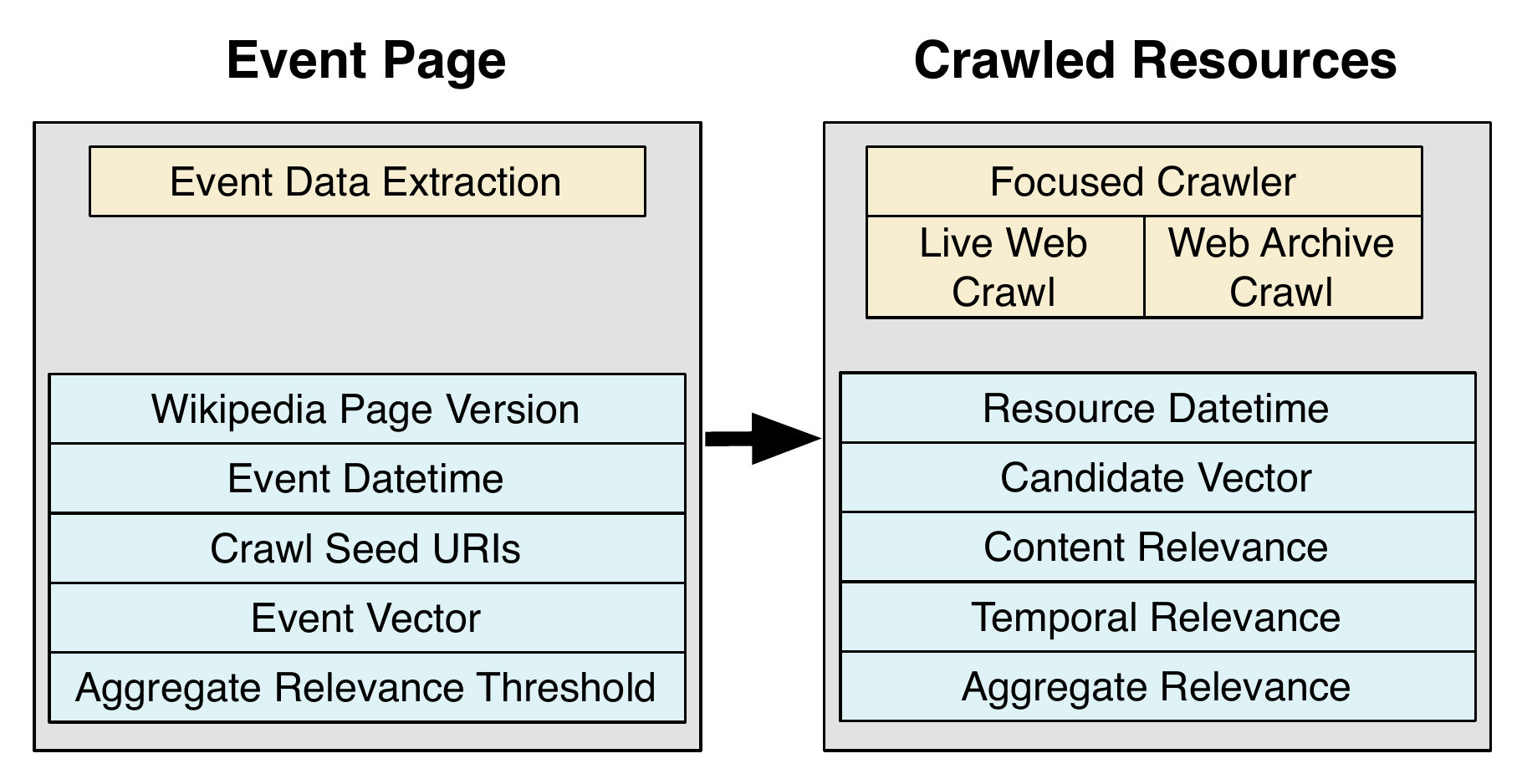}
\caption{Focused crawling framework}
\label{fig:framework}
\end{figure}
\section{Related Work}
Previous work by Gossen et al. \cite{gossen:extracting} inspired this work. They developed a focused
extraction (not web crawling) system to create event-centric collections from a large static archival collection
stored on a server under their control. The content of the Wikipedia page for an event is used to guide
the focused extraction. The event datetime is derived from HTML elements in the Wikipedia page and external 
references in that page are used as seed URIs. They found that their approach outperforms a naive extraction 
process that is not guided by content and that an approach that combines content-wise and temporal relevance 
scores mostly performs best. Our approach builds on this work. We deploy a focused crawler that operates on 
the real web and is not bound to a static, locally stored archival collection. We actually utilize $22$ web 
archives for our crawls and compare the results to comparable focused crawls on the live web.
%
%
A significant amount of work has been done on focused crawling in 
general \cite{CHAKRABARTI:focused_crawling,pant:crawl,aggarwal:crawling}.
Some work has additionally explored time-aware focused crawling, such as Pereira et al. \cite{pereira:crawling}.
In that work, the authors incorporated temporal data extracted from web pages to satisfy a particular
temporal focus of the crawl. They used temporal segmentation of text in a page to determine temporal focus.
We follow common practice for our focused crawling approach, for example, by implementing a priority
queue. The temporal segmentation of text could have been of interest for our temporal relevance
assessment, but, for this experiment we use extraction methods as seen in \cite{Farag2018}.
Relevant with regard to event-centric collection building is the work by Farag et al. \cite{Farag2018} and 
Littman et al. \cite{Littman2018}. Farag et al. introduced an intelligent focused crawling system that works 
on the basis of an event model. This model captures key data about the event and the focused crawler leverages 
the event model to predict web page relevance. As a result, the system can successfully filter unrelated content 
and perform at a high level of precision and recall. The work by Littmann et al. pertains to deriving event 
collections from social media. The authors focused on increasing the alignment between web archiving tools and 
processes, and social media data collection practice, for the overall goal of event-centric collection building.
Both efforts relate to our work in that the common goal is to build specific collections of web resources.
However, both Farag et al. and Littmann et al. are concerned with live web resources only.
\section{Establishing a Crawling Framework}
Our intent is to compare focused crawling of the live web and of web archives for the creation of collections pertaining
to unpredictable events such as natural disasters and mass shootings. Inspired by \cite{gossen:extracting}, we use the 
Wikipedia page that describes an event as a starting point. However, we do not use the current version of that page but rather
a prior version that is expected to describe the actual event and does not yet include post-event auxiliary content such as 
references to future related events or analysis of a range of similar events. We select external references of the Wikipedia 
version page as seeds for crawling and the page's text to assess content relevance of crawled resources. We additionally use 
a temporal interval starting with the datetime of the event to assess the temporal relevance of crawled resources. For both 
the live web and web archive crawls, crawled pages that are relevant, both content-wise and temporally, are added to the 
respective event collection. We describe the details in the remainder of this section and provide a conceptual overview of the 
framework in Figure \ref{fig:framework}.
\subsection{Wikipedia Page Version}
All data required to guide the crawling process is generated from the canonical Wikipedia page of the event. However, our 
events of interest have happened at some point in the past and their Wikipedia pages, very likely created shortly after 
the event, have with high probability evolved significantly since then.
This raises the question of which version of a Wikipedia page to use as the starting point for our crawls. Since Wikipedia
maintains all page versions along with the datetime they were created, we can, in theory, choose any version
between the very first and the current one. We know from related work \cite{Ratkiewicz:2010}
that the majority of edits to a Wikipedia page happen early on in its lifetime. However, event coverage often evolves beyond 
that point and hence consecutive page edits may still lead to significant changes. For example, other, related events may
happen at some later point and may result in the inclusion of new links and references into the event's Wikipedia page.
We therefore conjecture that using the current live version of an event's Wikipedia page could introduce too much content and 
references that do not directly pertain to a description of the event.
\begin{figure}[t]
\includegraphics[scale=0.25]{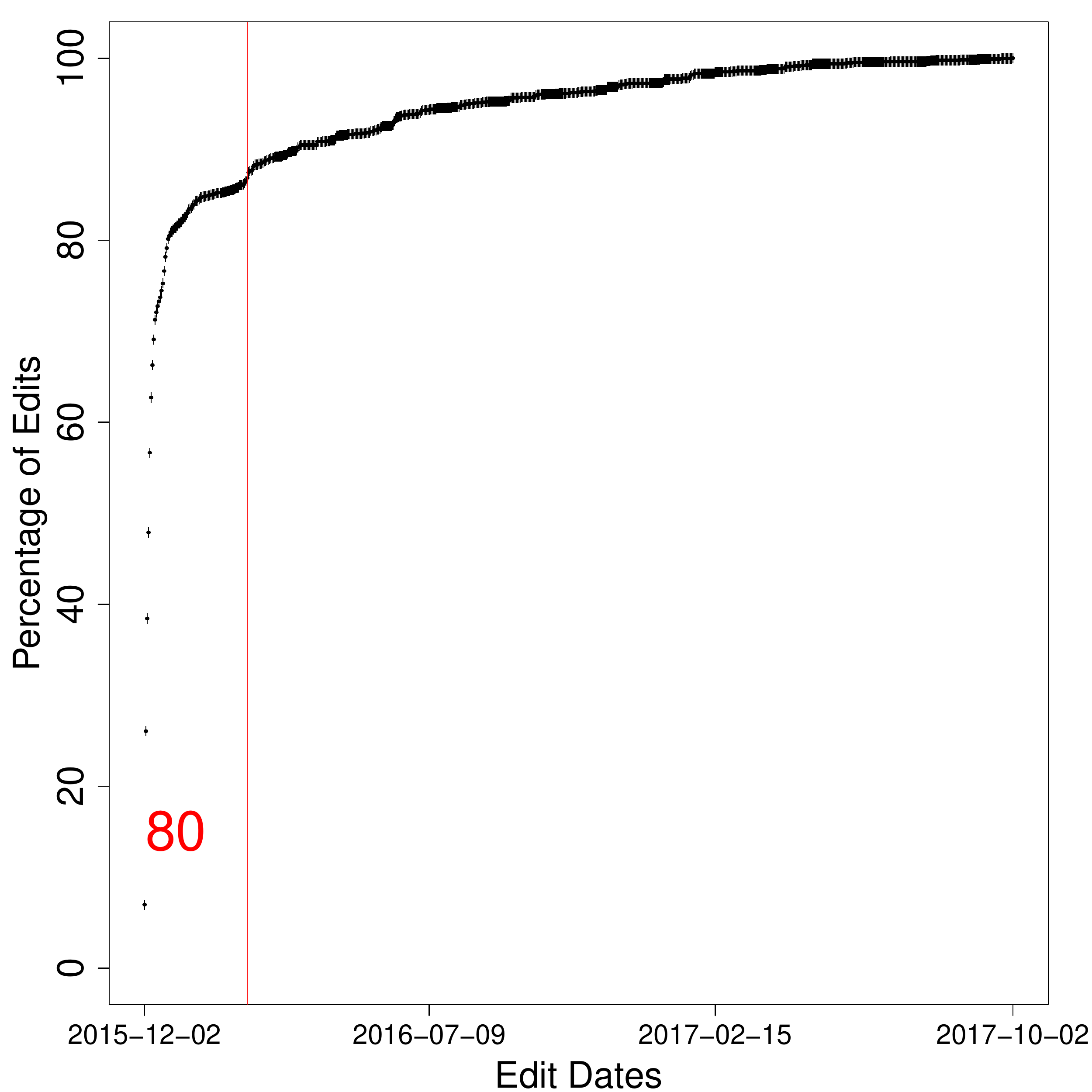}
\caption{Change point in Wikipedia article edit frequency}
\label{fig:change_point}
\end{figure}

We approach the selection of a Wikipedia page version from the perspective of edit frequency. Our goal is to
determine the date on which the vast majority of edits over the entire history of the page were completed. We select 
the page with that version date and consider that page to comprehensively capture the essence of the event.
For this purpose we plot all edits of a Wikipedia page and their datetimes. Figure \ref{fig:change_point} shows an example of such
a plot where the edit datetimes are on the x-axis and the percentage of edits on the y-axis. We then use the
standard $R$ changepoint library \cite{changepoint} to determine the change point in this graph. The change point is the
point after which the graph assumes a significantly different shape. In our case, this point is the datetime after which
the edit frequency drastically decreases. Hence, we can consider the page version that corresponds with that datetime 
as capturing the essence of the event. We refer to the change point datetime as $DT_{CP}$.
Figure \ref{fig:change_point} shows the edits of the San Bernadino Attack Wikipedia
page\footnote{\url{https://en.wikipedia.org/wiki/2015_San_Bernardino_attack}} and the detected change point at
$80$ days after the creation of the page. In this example, we select the version of the Wikipedia page that was live $80$ days
after the event\footnote{\url{https://en.wikipedia.org/w/index.php?title=2015_San_Bernardino_attack&oldid=706012350}} for our
experiment and refer to this version as the $DT_{CP}$ version of the Wikipedia page.
\subsection{Event Datetime}
The first data point that we extract from the $DT_{CP}$ version of the Wikipedia page is the event datetime. The 
format and granularity of the provided datetime can vary across Wikipedia pages. For uniformity, we express the
event datetime in date, month, year, hour, minute, and seconds. In case no exact time is available from the $DT_{CP}$ version 
of the Wikipedia page, we set the time to 00:00:01 of the day of the event. We refer to the event datetime as $DT_{E}$.
\subsection{Crawl Seed URIs}
Similar to \cite{gossen:extracting}, we extract all external references contained in the $DT_{CP}$ version of the Wikipedia 
page and consider their URIs as seeds for the focused crawl. For simplicity, we filter out references that do not point to 
English language content or that point to resources in a representation other than HTML. All remaining references are used 
as seeds for both the web archive and live web crawls as well as for the content relevance computation outlined below.
\subsection{Content Relevance}
This section describes the process aimed at determining the extent to which a crawled resource is content-wise relevant for 
inclusion in the event collection.
\subsubsection{Event Vector}
We use the textual content of the $DT_{CP}$ version of the Wikipedia page to create an event vector that will serve as our
baseline to assess the content relevance of crawled pages. In an effort to stabilize the event vector, we
further incorporate the textual content of a random $60\%$ of outgoing references from the $DT_{CP}$ version of the Wikipedia 
page. In order to reduce noise, such as advertisements, we apply the common boilerpipe
library\footnote{\url{https://github.com/kohlschutter/boilerpipe}}, introduced in
Kohlsch\"{u}tter et al. \cite{Kohlschutter:boilerpipe}, to the Wikipedia page as well as to its outgoing references.
From the remaining text of the page, we extract 1-grams and 2-grams, store their term frequency (TF),
and extract their inverse document frequency (IDF) from the Google NGram dataset \cite{goldberg2013dataset}.
These 1-grams and 2-grams, along with their combined TF-IDF score, make up the event vector.
\subsubsection{Candidate Vector and Content Relevance of a Crawled Resource}
The textual content of a crawled page is used to generate a candidate vector. We create this candidate vector in a manner
very similar to the event vector. After crawling a candidate page, we apply the boilerpipe library and extract the
remaining textual content. We determine TF-IDF values from extracted 1-grams and 2-grams to build the candidate
 vector. We then compute the cosine similarity between the candidate vector and the event vector to obtain a 
content relevance score $R_{cont}$. The resulting cosine value is between $0$ and $1$ where a higher score indicates a greater
level of similarity and hence content relevance of the crawled page. The way in which the content relevance is determined is
identical for resources in live web and web archive crawls.
\subsubsection{Content Relevance Threshold}
We compute a content relevance threshold for an event on the assumption that resources referenced in the $DT_{CP}$ version of 
the Wikipedia page are relevant themselves. We therefore run the same vector computation process for the content of the 
$40\%$ of references that remain after the process of generating the event vector and compute the cosine similarity
between both vectors. We repeat this process $10$ times, each time with a different random set of $60\%$ of references for
the event vector and hence different remaining $40\%$ of references for comparison. The computed average of the $10$ obtained
cosine similarity scores serves as our content relevance threshold $TH_{cont}$ for the event.
\subsection{Temporal Relevance}
This section describes the process aimed at determining the extent to which a crawled resource is temporally relevant for 
inclusion in the event collection.
\subsubsection{Temporal Interval and Temporal Relevance of a Crawled Resource}
\begin{figure}[t]
\includegraphics[scale=0.6]{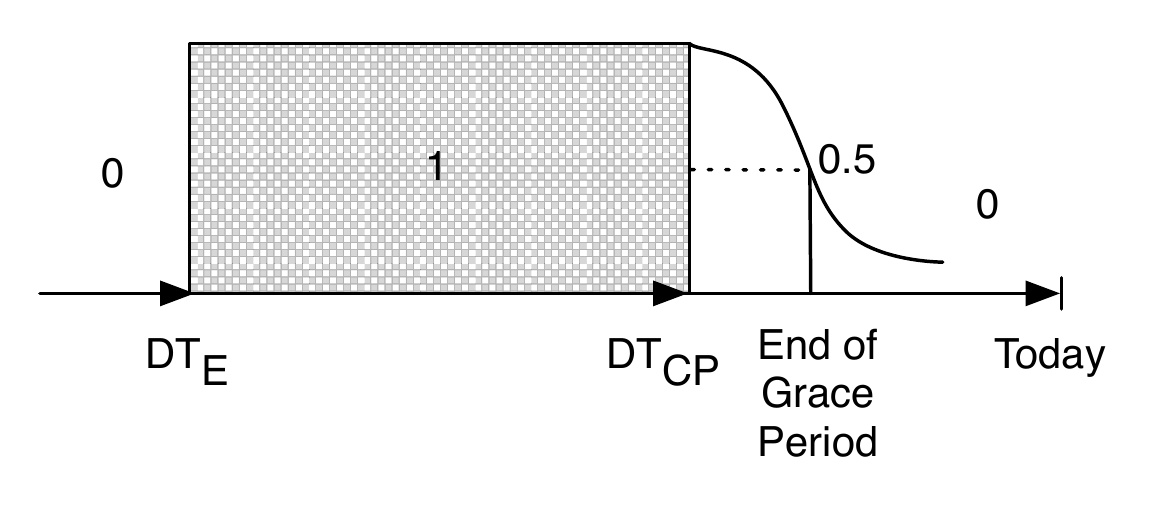}
\caption{Temporal Relevance Interval}
\label{fig:temporal_relevance}
\end{figure}
Inspired by \cite{gossen:extracting}, \cite{gossen:icrawl}, and \cite{Costa:tempranking}, we introduce a temporal interval 
to support assessing whether a crawled resource is temporally relevant. 
The interval, illustrated by Figure \ref{fig:temporal_relevance}, serves the purpose of assigning low temporal relevance
score to web resources that were published prior to the event or a long time after it.
Equation \ref{eq:rtemp} outlines the computation of the temporal relevance score, which we refer to as $R_{temp}$. For example, 
a crawled resource that has an associated datetime $DT_{R}$, for example its publication date (see \ref{subsubsec:datetime}),
prior to $DT_{E}$ gets a temporal relevance score of $R_{temp}=0$. A resource with $DT_{R}$ that falls between $DT_{E}$ and 
$DT_{CP}$, on the other hand, is assigned $R_{temp}=1$.
Additionally, a grace period beyond $DT_{CP}$ is considered. The grace period is not unlike the cool-down period introduced
in \cite{gossen:extracting} and is additionally motivated by the fact that web archives may take a while to archive a resource 
after it was published. For web archive crawls, the grace period provides a fair chance for resources that were published some 
time before $DT_{CP}$ but archived beyond it to still be considered relevant.
As can be seen in Equation \ref{eq:rtemp}, during the grace period, a resource can obtain a $R_{temp}$ score
of less than $1$ and greater or equal to $0.5$. In this equation,
${\Delta t\prime}$ represents the difference between $DT_{CP}$ and $DT_{R}$ and $\Delta t$ is equal to $1/4$ of the period
between $DT_{E}$ and $DT_{CP}$.
Different arguments can be made regarding the choice of the length of the grace period. Rather than setting a duration 
arbitrarily, we determine it using the time it took for references in the $DT_{CP}$ version of the Wikipedia page to 
be archived. More specifically, we use the average time between the datetimes associated with all references of the 
$DT_{CP}$ version of the Wikipedia page (as indicated in the article) and their corresponding archival datetime as the 
length of the grace period.
For resources captured in the live crawl, we apply a grace period to give certain resources published past $DT_{CP}$ a 
fair chance to be considered relevant.
In this case, we determine its duration as the average distance between the associated datetimes of all references from the 
$DT_{CP}$ version of the Wikipedia page (as indicated in the article).
\begin{equation}
R_{temp}~=~ 
\begin{cases}
1 & \quad \text{if } DT_{E} \le DT_{R} \le DT_{CP}\\
0 & \quad \text{if } DT_{E} > DT_{R} \\
e^{-\big( \left( \frac{ ln(2) } { \Delta t} \right) ~*~{\Delta t\prime}\big)} & \quad \text{if } DT_{R} > DT_{CP}\\
\end{cases}
\label{eq:rtemp}
\end{equation}
\subsubsection{Resource Datetime} \label{subsubsec:datetime}
As described in the previous section, the datetime $DT_{R}$ associated with a crawled resource plays a core role in determining 
its temporal relevance score. The manner in which this datetime is obtained is different for live web resources and Mementos.
To determine the $DT_{R}$ for a resource from the live web crawl, we use various approaches, some of which have also been used 
in Farag et al. \cite{Farag2018}. The first approach is to extract a datetime from the URI of a page, as many news publishers 
use URI patterns that contain a datetime, for example: \url{http://www.cnn.com/2017/12/09/us/wildfire-fighting-tactics/}. 
Second, we consider the page's HTML, as news publishers and content management systems frequently embed datetimes. For example, 
the following HTML excerpt is from a New York Times article:
\begin{verbatim}
<meta property="article:published" 
      itemprop="datePublished" 
      content="2017-12-09T10:14:50-05:00"/>
\end{verbatim}
Third, we utilize the CarbonDate tool\footnote{\url{http://carbondate.cs.odu.edu/}}, first introduced by SalahEldeen and 
Nelson \cite{carbondate}. The tool looks for first mentions of the URI on Twitter and Bitly. 
If these methods return more than one extracted datetime, we choose the earliest one as the page's $DT_{R}$.
If all methods fail and no datetime is extracted, we dismiss the crawled resource.

To determine the $DT_{R}$ for a resource from the archived web crawl, a feature of the Memento protocol \cite{memento}
that is supported by all archives included in the experiment, is leveraged because it yields a datetime with minimal 
effort involved. A web archive that returns a Memento also returns the datetime it was archived in the 
\texttt{Memento-Datetime} HTTP response header. If this datetime falls within the temporal interval between $DT_{E}$ and 
$DT_{CP}$ as shown in Figure \ref{fig:temporal_relevance}, we use it as our $DT_{R}$. This kind of Memento will obtain a 
temporal relevance score of $1$.
Understanding that web archives commonly archive pages quite some time after they were published,
this approach can lead to pages that were published prior to $DT_{E}$ (but archived past it) receiving a score of $1$.
However, given the temporal threshold will be combined with a content-based threshold,
this risk is outweighed by the benefit of a straightforward means to determine a $DT_{R}$.
In cases where the archival datetime is beyond $DT_{CP}$, we can not merely dismiss the Memento because it could have 
been archived a long time after it was initially published. Hence, in these cases we attempt to determine the publication 
date of the page on the basis of the Memento. To that end, we use the CarbonDate tool again. If the tool can assign a date 
to the Memento, we use it as $DT_{R}$. If the tool is unsuccessful, we leverage archived HTTP headers, which some web 
archives convey as custom X-headers in the HTTP response of a Memento. For example, if a Memento provides an 
\texttt{X-Last-Modified} header, we use its datetime as $DT_{R}$. If all methods fail, the crawled resource is dismissed.
\begin{figure*}[th!]
    \centering
    \begin{subfigure}[t]{0.3\textwidth}
        \centering
        \includegraphics[scale=0.1]{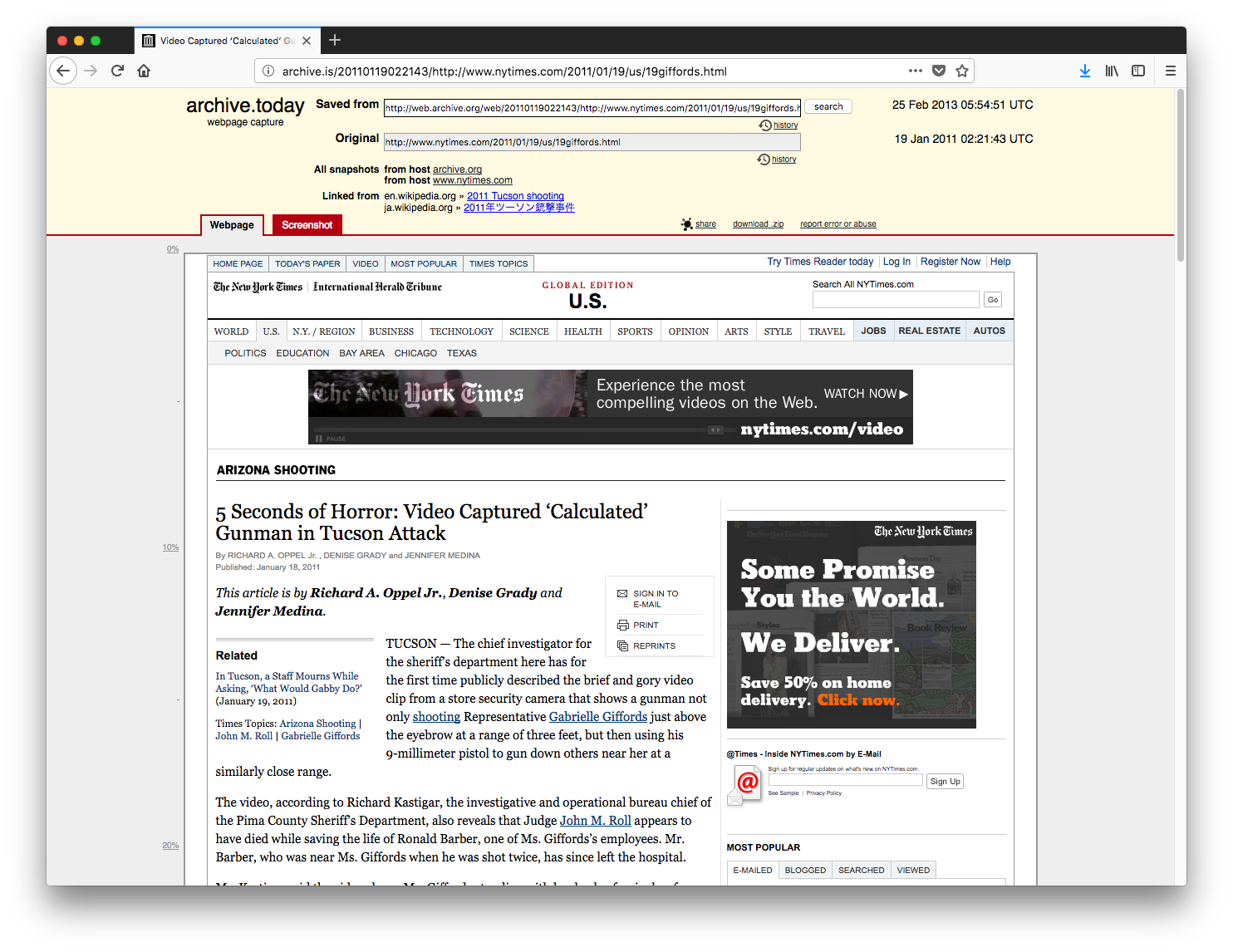}
        \caption{Depth 0, archive.today, $R_{aggr}=0.89$} 
        \label{fig:lev0}
    \end{subfigure}
    ~
    \begin{subfigure}[t]{0.3\textwidth}
        \centering
        \includegraphics[scale=0.1]{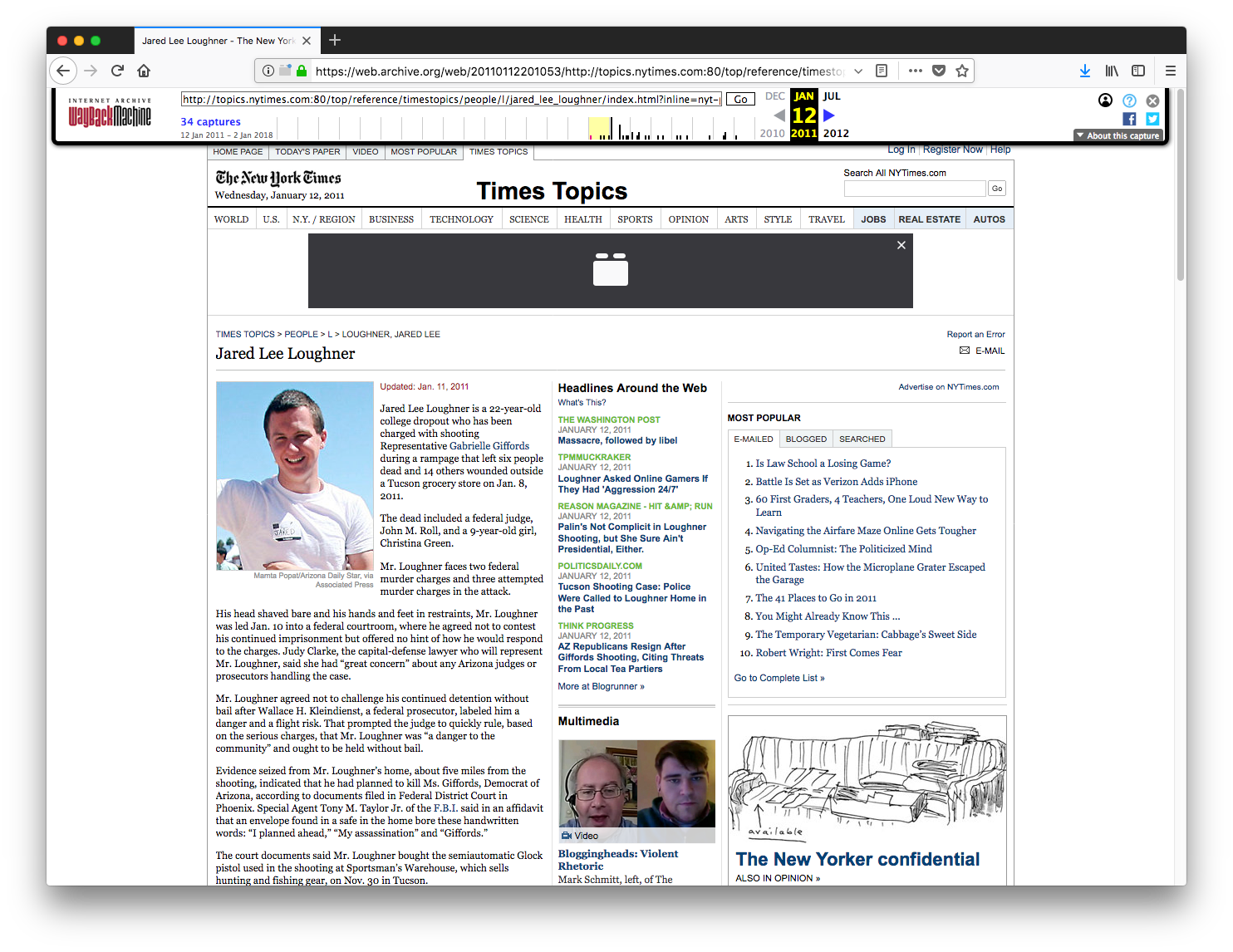}
        \caption{Depth 1, Internet Archive, $R_{aggr}=0.90$} 
        \label{fig:lev1}
    \end{subfigure}
    ~
    \begin{subfigure}[t]{0.3\textwidth}
        \centering
        \includegraphics[scale=0.1]{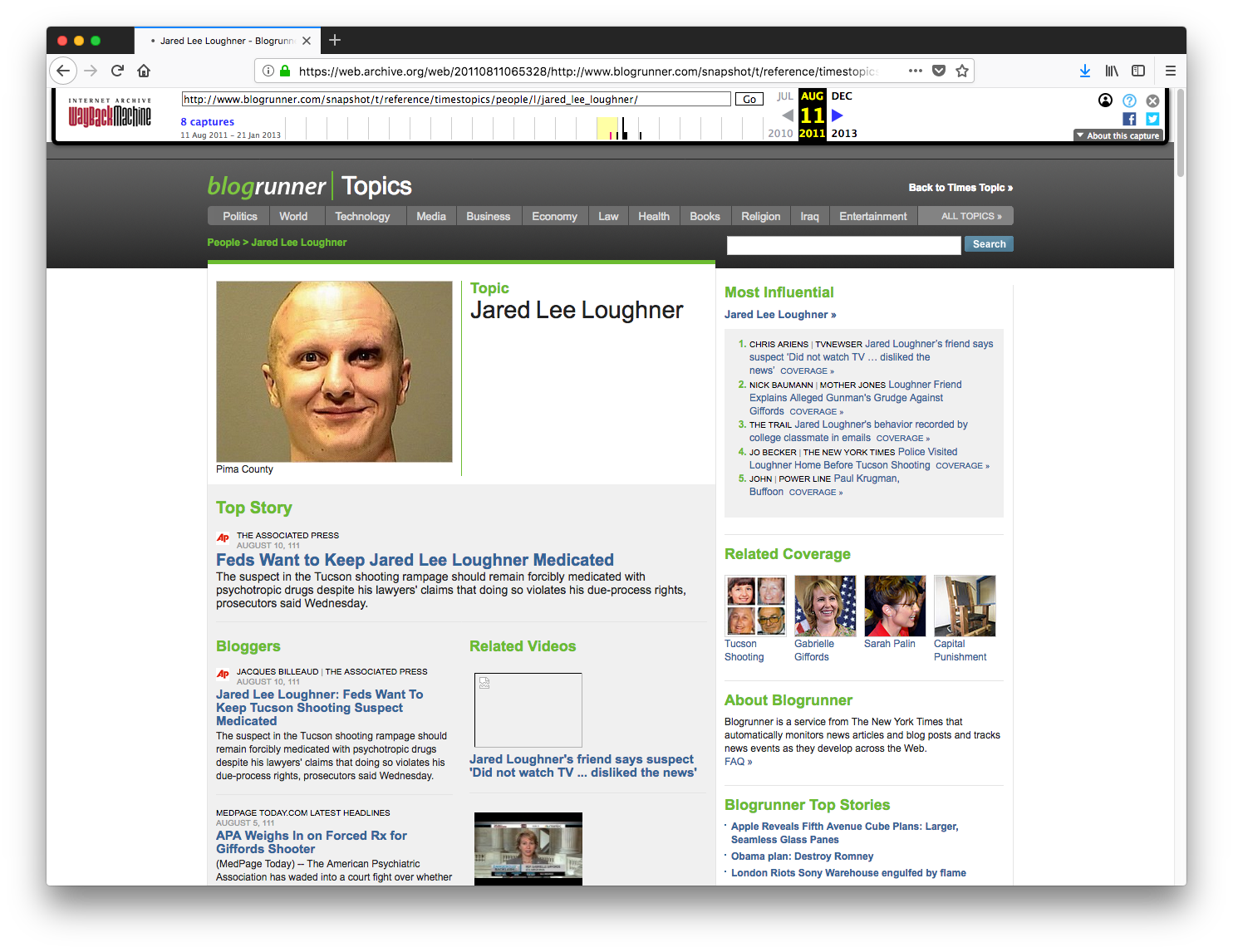}
        \caption{Depth 2, Internet Archive, $R_{aggr}=0.89$} 
        \label{fig:lev2}
    \end{subfigure}
    \begin{subfigure}[t]{0.3\textwidth}
        \centering
        \includegraphics[scale=0.1]{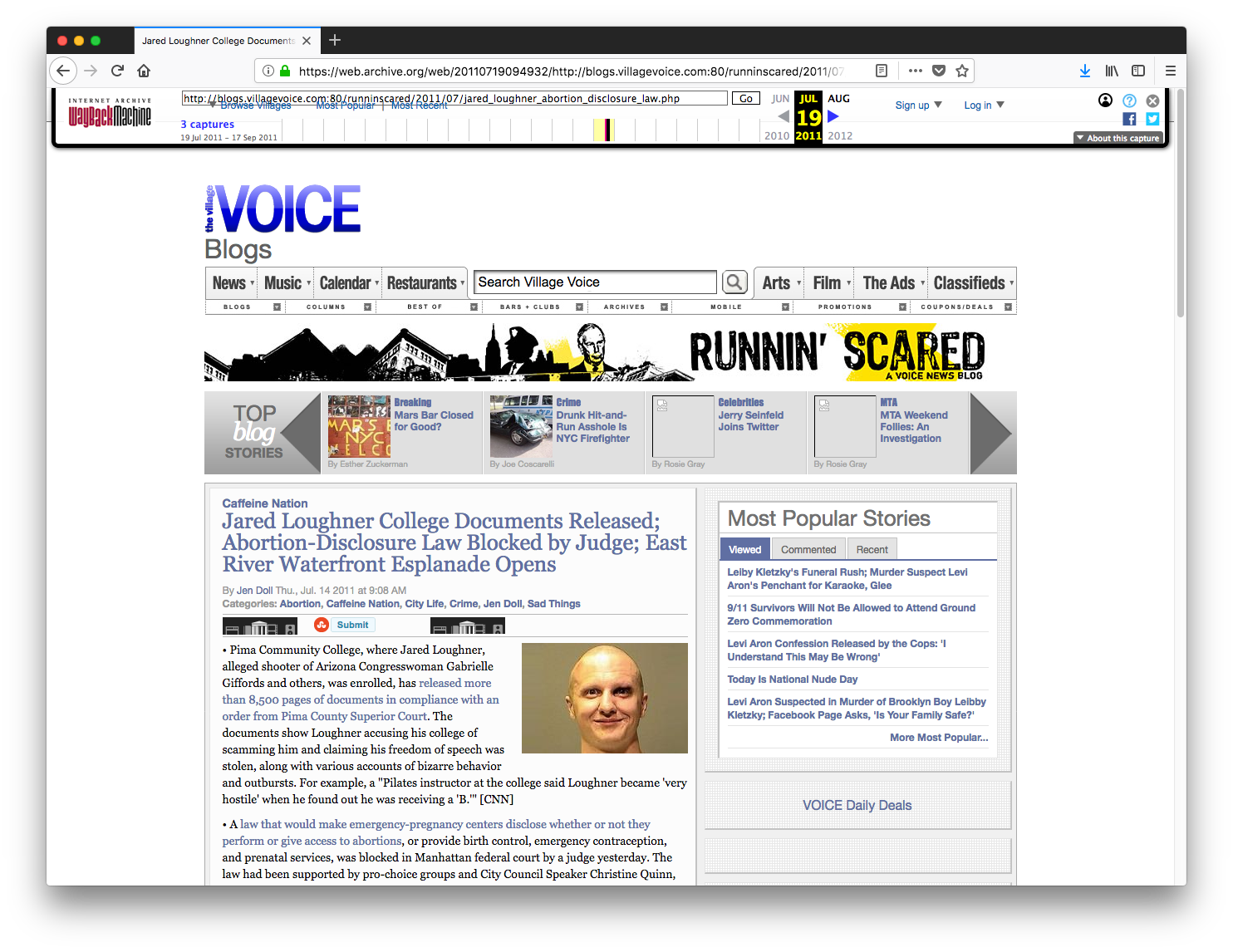}
        \caption{Depth 3, Internet Archive, $R_{aggr}=0.89$} 
        \label{fig:lev3}
    \end{subfigure}
    ~
    \begin{subfigure}[t]{0.3\textwidth}
        \centering
        \includegraphics[scale=0.1]{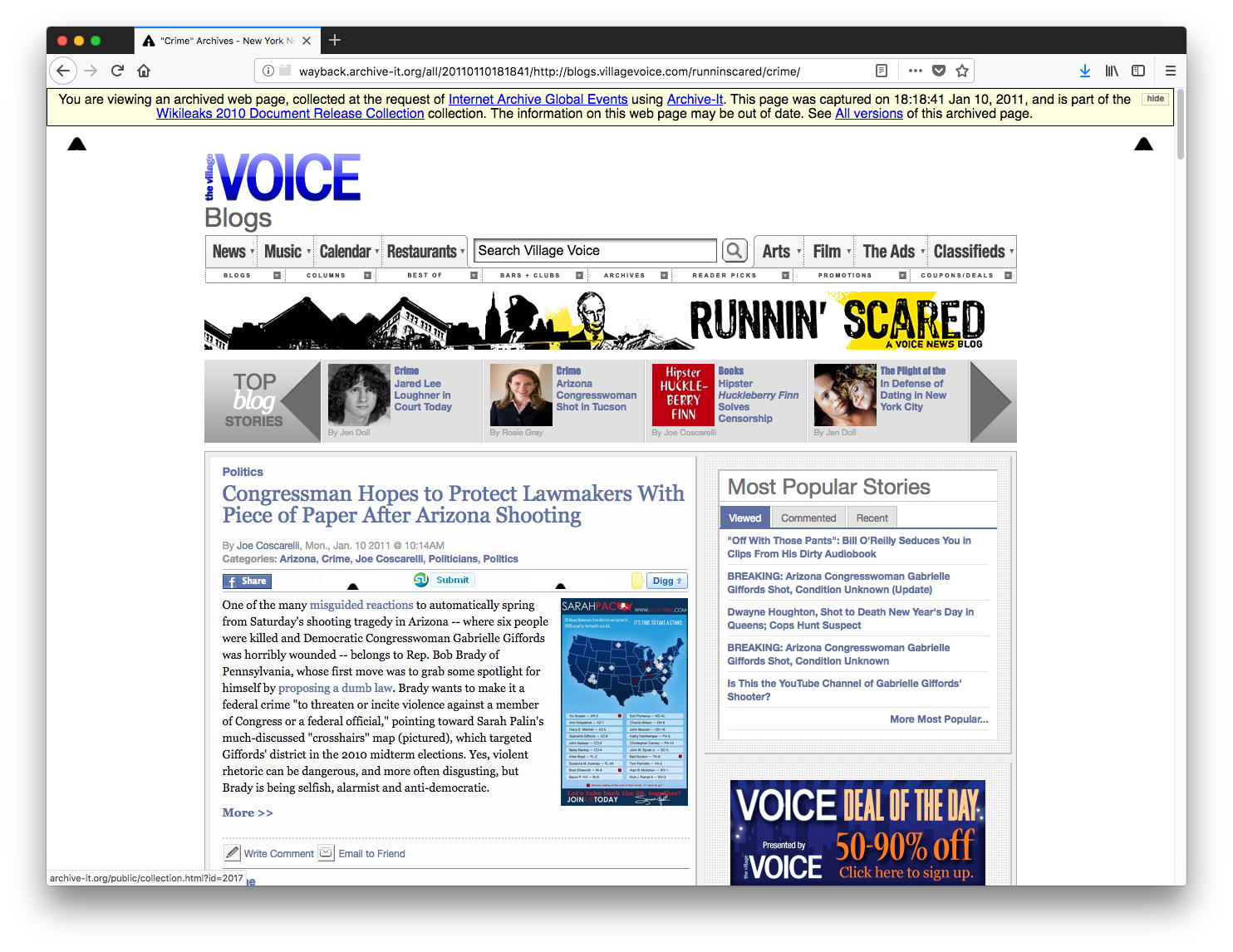}
        \caption{Depth 4, Archive-It, $R_{aggr}=0.91$} 
        \label{fig:lev4}
    \end{subfigure}
    ~
    \begin{subfigure}[t]{0.3\textwidth}
        \centering
        \includegraphics[scale=0.1]{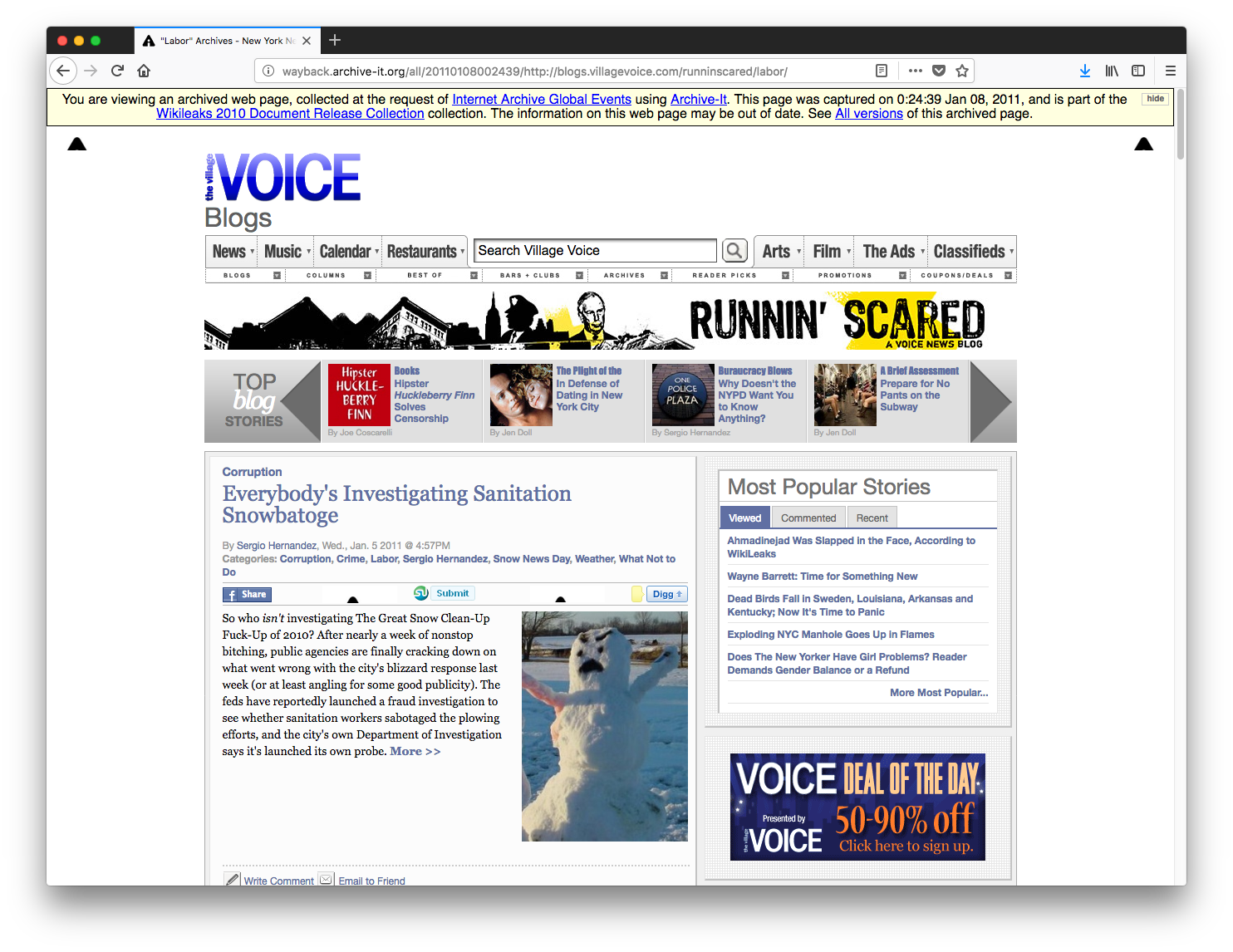}
        \caption{Depth 5, Archive-It, $R_{aggr}=0.51$} 
        \label{fig:lev5}
    \end{subfigure}
    \caption{Mementos resulting from the TUC web archive crawl at depth 0 (seed) through depth 5 obtained from various web archives using the Memento infrastructure}
    \label{fig:crawls}
\end{figure*}
\subsubsection{Temporal Relevance Threshold}
We compute a temporal relevance threshold for an event on the assumption that resources referenced in the $DT_{CP}$ version of 
the Wikipedia page are temporally relevant themselves. We therefore compute the temporal relevance of each URI in the same 
random set of $60\%$ of references that we use for the computation of $TH_{cont}$. We repeat this process $10$ times, each time 
with a different set of random $60\%$ and use the computed average of all obtained scores as our temporal relevance threshold 
$TH_{temp}$. 
\subsection{Aggregate Relevance and Aggregate Relevance Threshold}
Following the same reasoning as in \cite{gossen:extracting}, we use an aggregate relevance score $R_{aggr}$ based on the sum of the
content and temporal relevance scores, respectively $R_{cont}$ and $R_{temp}$.
In order to aggregate both scores into one, we introduce two weighting factors $\alpha$ and $\beta$, as shown in
Equation \ref{eq:r_aggr}. These factors can be used to weigh the significance of either relevance score. For our experiments
we balance the weight equally and assign the value of $0.5$ to both $\alpha$ and $\beta$, as also seen
in \cite{gossen:extracting}.
\begin{equation}
R_{aggr}~=~\alpha*R_{cont} + \beta*R_{temp}
\label{eq:r_aggr}
\end{equation}
\begin{equation}
TH_{aggr}~=~\alpha*TH_{cont} + \beta*TH_{temp}
\label{eq:th_aggr}
\end{equation}
Similarly, as shown in Equation \ref{eq:th_aggr}, we define an aggregate threshold. We 
use the same weighting factors as seen in Equation \ref{eq:r_aggr} to balance the significance of both parts.

Based on the $R_{aggr}$ score of a page and the computed $TH_{aggr}$ of the corresponding event, we determine whether the crawled
page will be selected for the event collection or not. We classify a page with an aggregate relevance score equal to or above the 
threshold ($R_{aggr} \ge TH_{aggr}$) as relevant and hence select it into the collection. On the other hand, we consider a page
with a score below the threshold ($R_{aggr} < TH_{aggr}$) as not relevant and reject it.
\begin{table*}[th!]
\caption{Crawled events}
\label{tab:events}
\begin{tabular}{|l|l|c|l|} \hline
\textbf{Event} & \boldmath{$DT_{E}$} & \boldmath{$DT_{CP}$} & \textbf{Wikipedia page version} \\ \hline \hline
NYC & 10/31/2017 & NA & \url{https://en.wikipedia.org/wiki/2017_New_York_City_truck_attack}\\ \hline
SB & 12/02/2015 & 02/20/2016 & \url{https://en.wikipedia.org/w/index.php?title=2015_San_Bernardino_attack&oldid=706012350} \\ \hline
TUC & 01/08/2011 & 01/12/2012 & \url{https://en.wikipedia.org/w/index.php?title=2011_Tucson_shooting&oldid=471037980} \\ \hline
BIN & 04/03/2009 & 11/11/2009 & \url{https://en.wikipedia.org/w/index.php?title=Binghamton_shootings&oldid=325176468}\\ \hline
\end{tabular}
\end{table*}
\section{Crawling the Live and Archived Web}
Our crawling process, just like other implementations of focused crawlers, is deployed with a priority queue that
informs the crawler which URIs to crawl next. In our case, resources linked from pages with a higher aggregate 
relevance score will be ranked higher in the priority queue. Our crawling process also needs to stop at some point.
The simplest stop condition for a focused crawler is when the queue is empty and there are no documents left to
crawl. However, under this condition, depending on the event and the length of the list of seed URIs, the crawl can run
for a long time. Other typical stop conditions for crawlers are a maximum number of documents crawled, a maximum size of
the crawled dataset, a maximum runtime, or a maximum crawl depth. We chose to implement the latter condition and run our
focused crawler for a maximum depth of six. A seed URI is considered depth $0$ and as long as the outlinks remain
relevant, our crawler follows outlinks up until crawl depth $5$. Arguably, the chosen crawl depth is somewhat arbitrary
but our preliminary tests indicated that smaller depths tended to result in too
few documents and larger depths took too long to complete. Clearly, this stop condition is
configurable and we leave a thorough investigation of an optimal stop condition for future work.
We modify the code base of the \texttt{crawler4j}\footnote{\url{https://github.com/yasserg/crawler4j}} tool for our focused 
crawler and run all crawls on an Amazon virtual machine.

The remainder of this section provides further details about the crawling process with a focus on web archive crawling 
because, to the best of our understanding, the work described here is the first to use focused crawling across web archives.
\subsection{Live Web Crawls}
The crawl of the live web follows established focused crawl practice, starting by fetching a seed URI page from the live 
web and determining and evaluating its aggregate relevance $R_{aggr}$. If the page is deemed relevant, it is added to the event 
collection, its outlinks are extracted and added to the priority queue. Each URI in the priority queue is handled in the same 
manner until the crawler's stop condition is met.
\subsection{Web Archive Crawls} \label{subsec:archived_web_crawls}
Crawling the archived web is done by utilizing the Memento protocol \cite{memento} and associated infrastructure. Unlike 
previous work \cite{gossen:extracting}, in order to generate the richest possible event collections, we are interested in 
obtaining Mementos from as many publicly available web archives around the world as possible. 
The Memento infrastructure, and in particular the Memento Aggregator \cite{Bornand:MementoML}, makes this possible.
For each URI that needs to be crawled (seed URIs and URIs in the priority queue) until the crawler's stop condition is met,
the crawler obtains a Memento of that URI that was archived temporally closest but after $DT_{E}$. Closest to that datetime, 
in order to avoid using a version of the resource for which the content may have drifted \cite{jones:content_drift} since it was
originally linked to. And after that datetime because, clearly, pages that were archived prior to $DT_{E}$ were also published 
before it and hence are not relevant when unplanned events are concerned.
The Memento protocol and the Memento Aggregator provide two ways to discover a Memento with an archival date
closest to a desired date. The TimeMap approach consists of requesting a list of URIs of all available Mementos 
(URI-Ms in Memento protocol lingo) for a certain original URI (URI-R in Memento protocol lingo). From that list, the Memento 
closest to and after $DT_{E}$ can be selected. The TimeGate approach entails performing datetime negotiation by providing an 
original URI as well as a preferred archival datetime, and receiving the URI of the Memento with an archival datetime temporally 
closest to the preferred datetime in return. However, this approach can yield a Memento that is either prior to or after the 
preferred datetime. Both the TimeMap and TimeGate approaches require the Memento Aggregator to issue a request to multiple web 
archives for each URI. As such, in both cases, extra HTTP requests are involved when compared to live web crawling where a 
URI is accessed directly. Therefore, a web archive crawl will necessarily be slower than a live web crawl. However, the 
TimeMap approach can involve significantly more HTTP requests than the TimeGate approach because obtaining a complete TimeMap 
from a single archive itself may entail multiple requests. As such, in order to reduce the overall web archive crawling time, 
we use the TimeGate approach for our experiments and use $DT_{E}$ as the preferred datetime.
In case the returned Memento has an archival datetime prior to our $DT_{E}$, we simply follow the \texttt{next memento} HTTP 
link header, which is provided in the TimeGate HTTP response. This header points to the temporally ``next'' Memento that, as 
per the Memento protocol's datetime negotiation, has a datetime greater or equal to $DT_{E}$.

For each URI that needs to be crawled, this process yields the URI of a Memento. The crawler fetches that Memento from the web 
archive that holds it, computes its $R_{aggr}$ score and evaluates it vis-a-vis the $TH_{aggr}$. If the Memento is deemed 
relevant, it is added to the event collection, its outlinks are extracted and added to the priority queue. We note that most 
web archives rewrite outlinks in their Mementos to point back into the same archive rather than to the live web, even when 
the archive does not hold a Memento for the linked resource or only holds Mementos that are temporally distant from the desired 
time, which in our case is $DT_{E}$ \cite{ainsworth:evaluating}. We therefore add the original URI (URI-R) of the outlink,
which can be obtained using features of the Memento protocol, to the priority queue rather than the rewritten URI-M of the 
outlink. This allows us to discover the Memento for outlinks that is temporally closest and past the event datetime $DT_{E}$ 
across all web archives covered by the Aggregator.

Figure \ref{fig:crawls} shows six screenshots of consecutively crawled Mementos. Figure \ref{fig:lev0} shows the Memento of
the seed URI, Figure \ref{fig:lev1} the Memento of one of the seed's outlinks (crawl depth $1$), Figure \ref{fig:lev2} the Memento 
of an outlink of the prior Memento of crawl depth $1$ (crawl depth $2$), and so on. These screenshots show the diversity of 
contributing web archives: the Memento for the seed URI was found in \texttt{archive.today}, the Mementos for crawl depths 
$1..3$ were provided by the Internet Archive, and depths $4..5$ by Archive-It. The figure also shows the $R_{aggr}$ scores for 
each Memento. The threshold $TH_{aggr}$ for this crawl was $0.75$ and hence the first five Mementos
are classified as relevant but the last one is not. Since our crawl depth was set to five, the outlinks of the Memento shown 
in Figure \ref{fig:lev5} were not added to the priority queue. If, however, it had been set to a number larger than five, the 
Memento's outlinks would also not have been added to the queue as the Memento's $R_{aggr} < TH_{aggr}$.
\begin{figure*}[t]
    \centering
    \begin{subfigure}[t]{0.49\textwidth}
        \centering
        \includegraphics[scale=0.166]{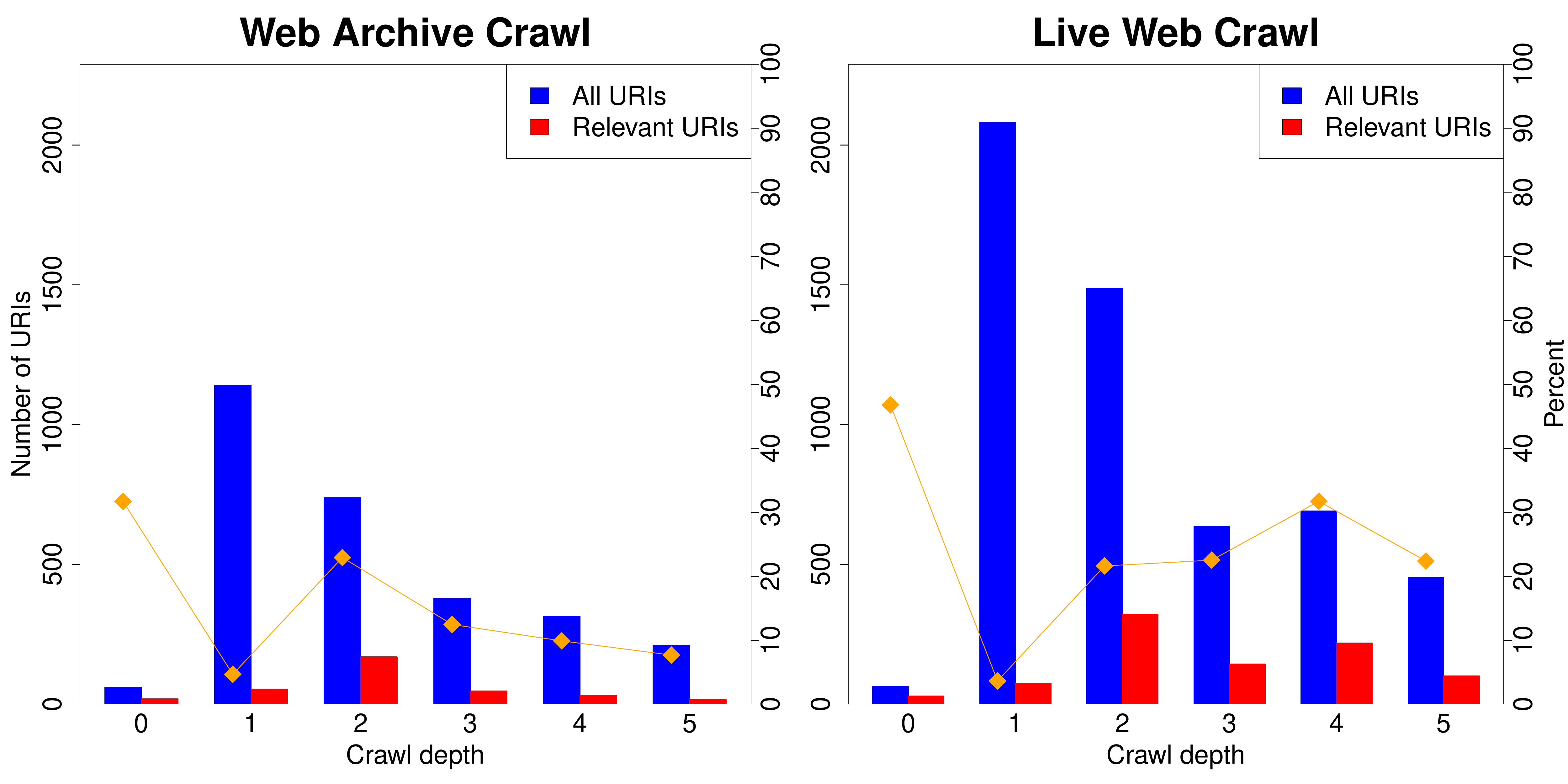}
        \caption{New York City}
        \label{fig:nyc_rel_uris}
    \end{subfigure}
    ~
    \begin{subfigure}[t]{0.49\textwidth}
        \centering
        \includegraphics[scale=0.166]{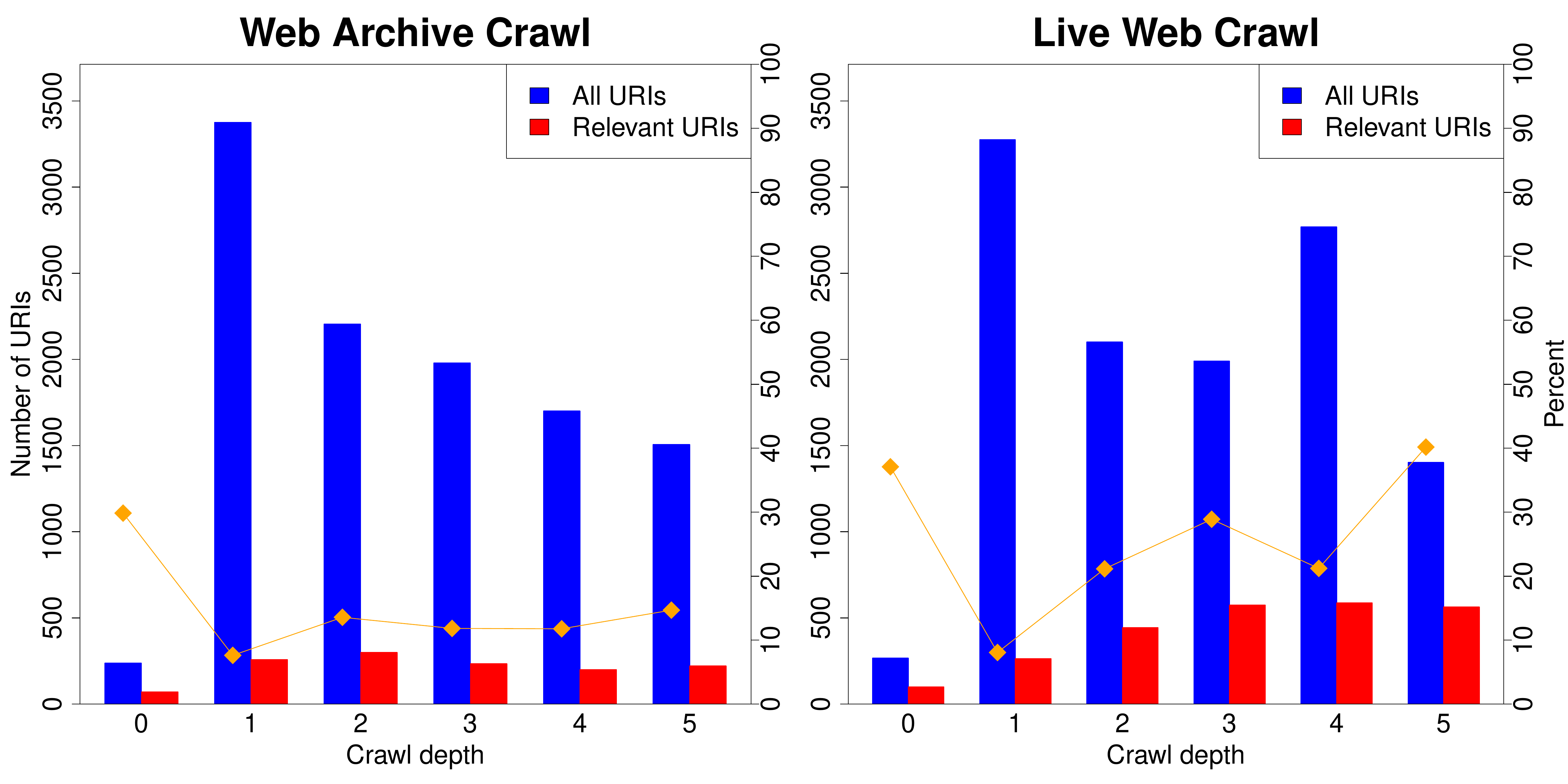}
        \caption{San Bernadino}
        \label{fig:sb_rel_uris}
    \end{subfigure}
    \begin{subfigure}[t]{0.49\textwidth}
        \centering
        \includegraphics[scale=0.166]{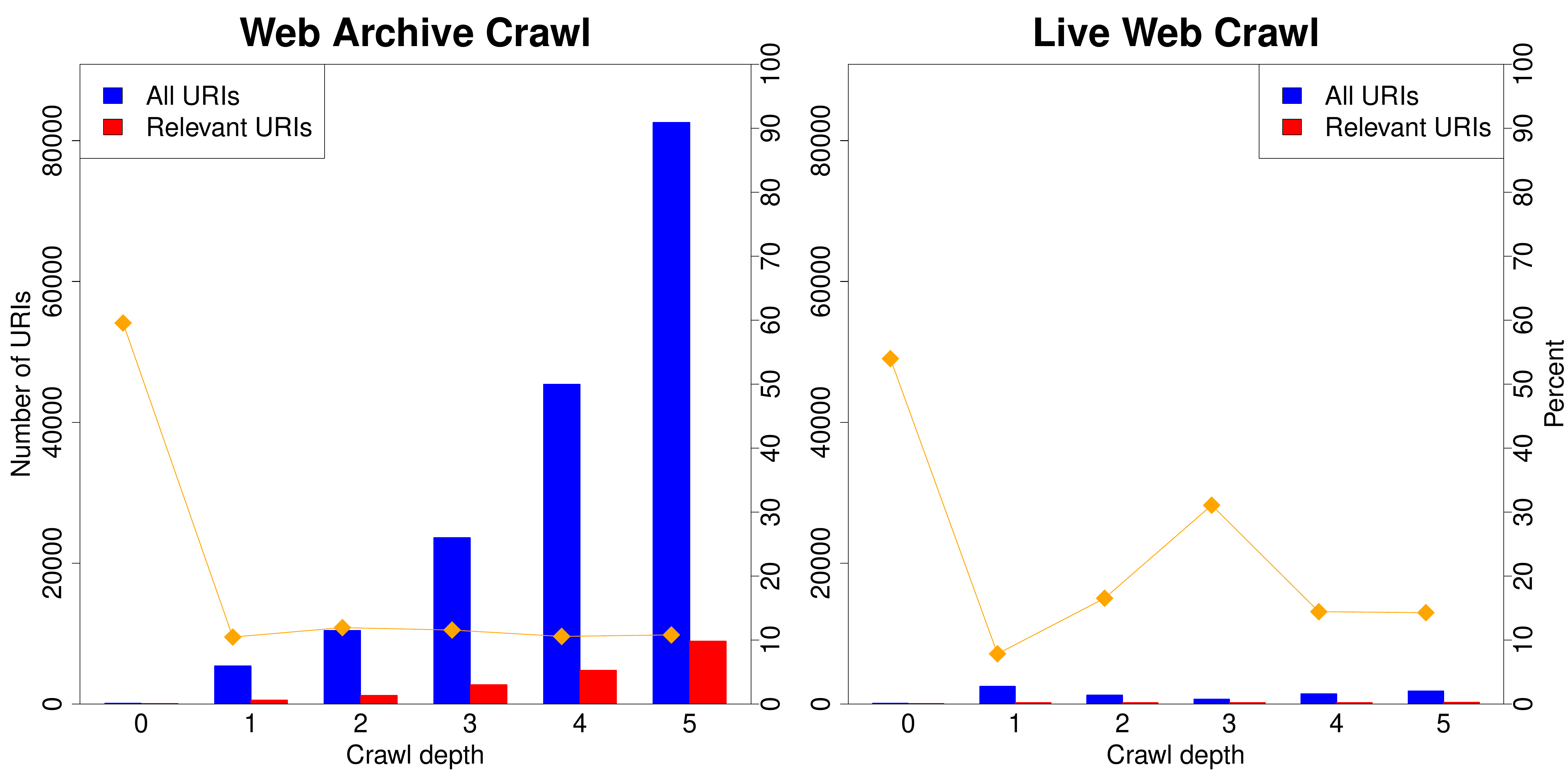}
        \caption{Tucson}
        \label{fig:tu_rel_uris}
    \end{subfigure}
    ~
    \begin{subfigure}[t]{0.49\textwidth}
        \centering
        \includegraphics[scale=0.166]{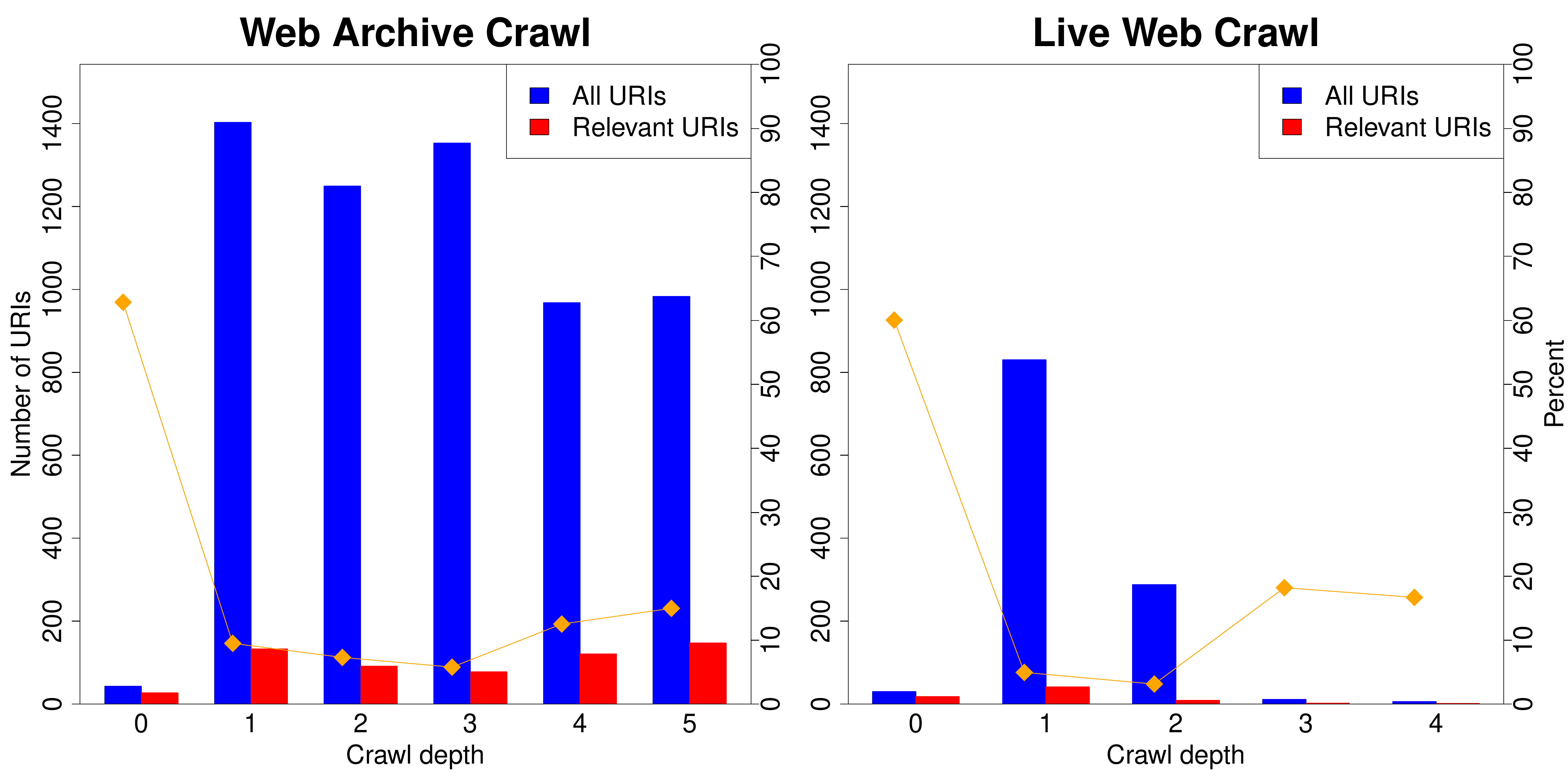}
        \caption{Binghampton}
        \label{fig:bi_rel_uris}
    \end{subfigure}
    \caption{Relevant URIs}
    \label{fig:rel_uris}
\end{figure*}
\section{Web Crawl Comparison}
We present the results of crawls for four different events: the 2017 New York City attack (NYC), the 2015 San
Bernadino attack (SB), the 2011 Tucson shooting (TUC), and the 2009 Binghampton shootings (BIN).
We chose these events because they are fairly similar in nature, they all happened in the U.S., and
their coverage on the web is predominantly in English. We assumed that this uniformity would better support
detecting patterns in our results.
We ran our crawls in November of 2017, a few days after the New York City attack, and more than eight
years after the Binghampton shootings. Table \ref{tab:events} summarizes the four events for which we created
an event collection with our focused crawling framework. The table also shows the event dates $DT_{E}$,
the change points $DT_{CP}$, and the URIs of the $DT_{CP}$ versions of the Wikipedia event page.
Note that we did not compute a change point for the NYC event because we crawled resources very soon after 
the attack happened, at which point the number of Wikipedia page edits had not yet reached the change point.
As such, for the NYC event, we used the live version of the Wikipedia event page as it was at the time of crawling.
\subsection{Relevant URIs}
Our first results are visualized in Figure \ref{fig:rel_uris}, distinguished by event. For example, the crawl data for
the New York City attack is shown in Figure \ref{fig:nyc_rel_uris},
for the San Bernadino attack in Figure \ref{fig:sb_rel_uris},
and so on. The left-hand plot for each event shows the results from the web archive crawl,
and the right plot displays our results from the live web crawl.
All subfigures of Figure \ref{fig:rel_uris} show the number of URIs crawled at each crawl depth (0..5).
The blue bars indicate
the total number of URIs crawled and the red bars represent the number of URIs that were classified as relevant,
per corresponding crawl depth. The bars refer to the left y-axis.
The lines, representing the fraction of relevant URIs, refer to the right y-axis.
For the NYC event, the live web crawl is the clear winner as it returns significantly more URIs as well as relevant URIs.
The fraction of relevant URIs on depth $0$ (the seeds) is almost $50\%$ for the live web vs. $30\%$ for the web
archive crawl. On crawl depth $1$, the first outlinks from the seeds, and on depth $2$,
the fractions are fairly similar. But for the further depths $3$, $4$, and $5$ the live crawl shows
ratios above $20\%$ of relevant URIs whereas the web archive crawl only shows ratios around $10\%$.
This result makes intuitive sense as we conducted the crawl merely days after the event happened.
It is highly likely that web archives did not have a chance to archive a significant amount of
the relevant resources and hence our web archive crawl did not surface many (relevant) URIs.
The results for the SB crawls, shown in Figure \ref{fig:sb_rel_uris}, are similar in that the live crawl returns a 
higher ratio of relevant URIs at all crawl depths. While the number of total URIs crawled is comparable between 
both crawls, the number of relevant URIs is consistently higher for the live web crawl. Our interpretation of these 
results is that, since the event datetime is two years in the past, web archives have had enough time to create 
Mementos of many relevant web pages. However the web archive crawl does not outperform the live web crawl.
Figures \ref{fig:tu_rel_uris} and \ref{fig:bi_rel_uris} show a very different pattern.
In both cases the live web crawl results in
fewer total URIs and fewer relevant URIs crawled than the web archive crawl.
The BIN live crawl does not even return any URIs on depth $5$.
Our interpretation of this pattern is based on the fact that the TUC and BIN events happened in 2011 and 2009,
respectively. Hence, a lot of time has passed for pages on the live web to either completely disappear
or to have their content drift to something less relevant compared to the event vector.
This is a phenomenon that we have previously investigated
in the realm of scholarly communication \cite{klein:one_in_five,jones:content_drift} and that seems to also happen
for web coverage of unplanned events. In essence, our finding suggests that live web resources pertaining to an event 
that were available at the time of the event are by now more likely available in web archives than on the live web.
\begin{figure*}[t!]
    \centering
    \begin{subfigure}[t]{0.49\textwidth}
        \centering
        \includegraphics[scale=0.166]{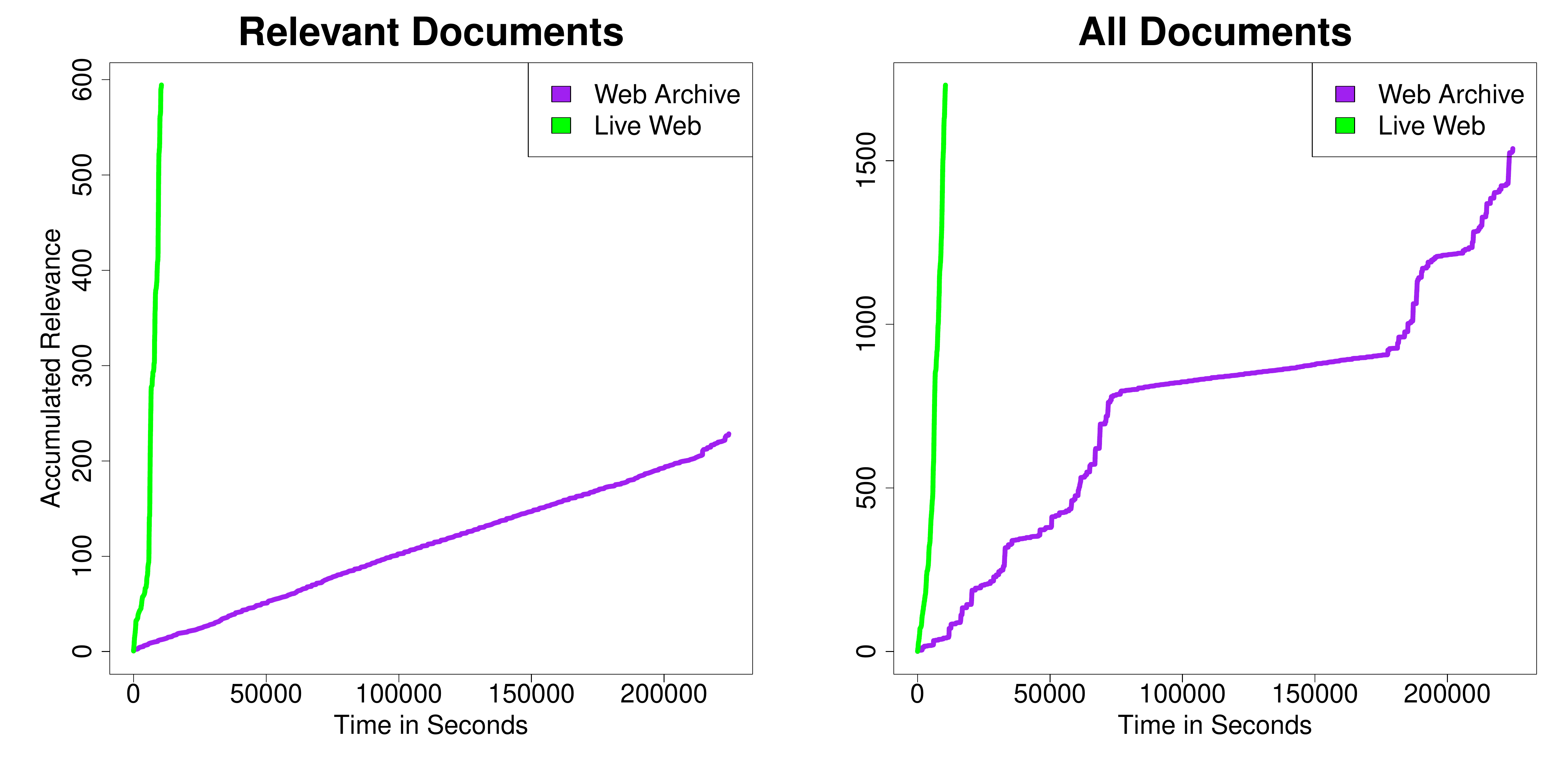}
        \caption{New York City}
        \label{fig:nyc_rel_time}
    \end{subfigure}
    ~
    \begin{subfigure}[t]{0.49\textwidth}
        \centering
        \includegraphics[scale=0.166]{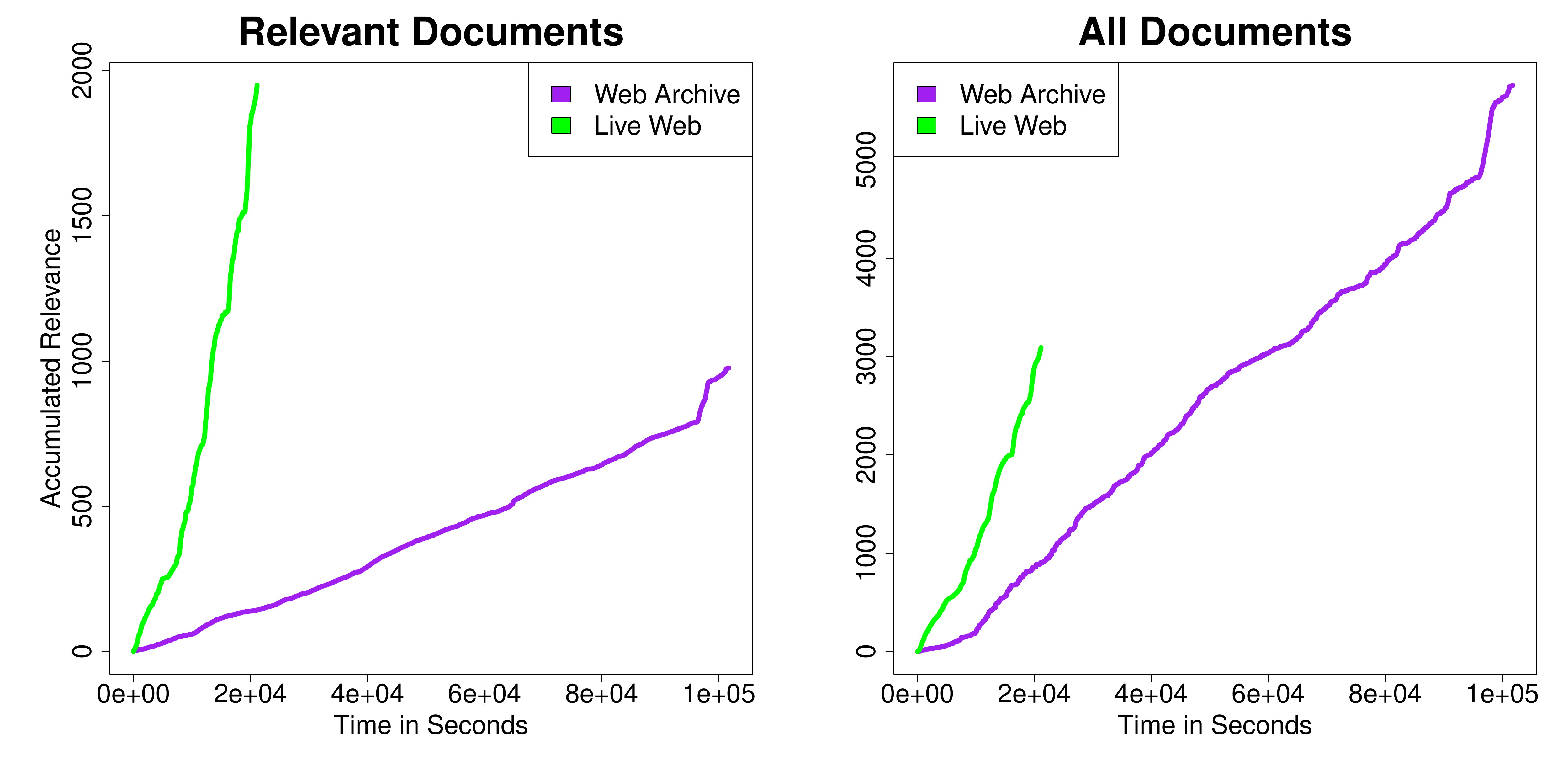}
        \caption{San Bernadino}
        \label{fig:sb_rel_time}
    \end{subfigure}
    \begin{subfigure}[t]{0.49\textwidth}
        \centering
        \includegraphics[scale=0.166]{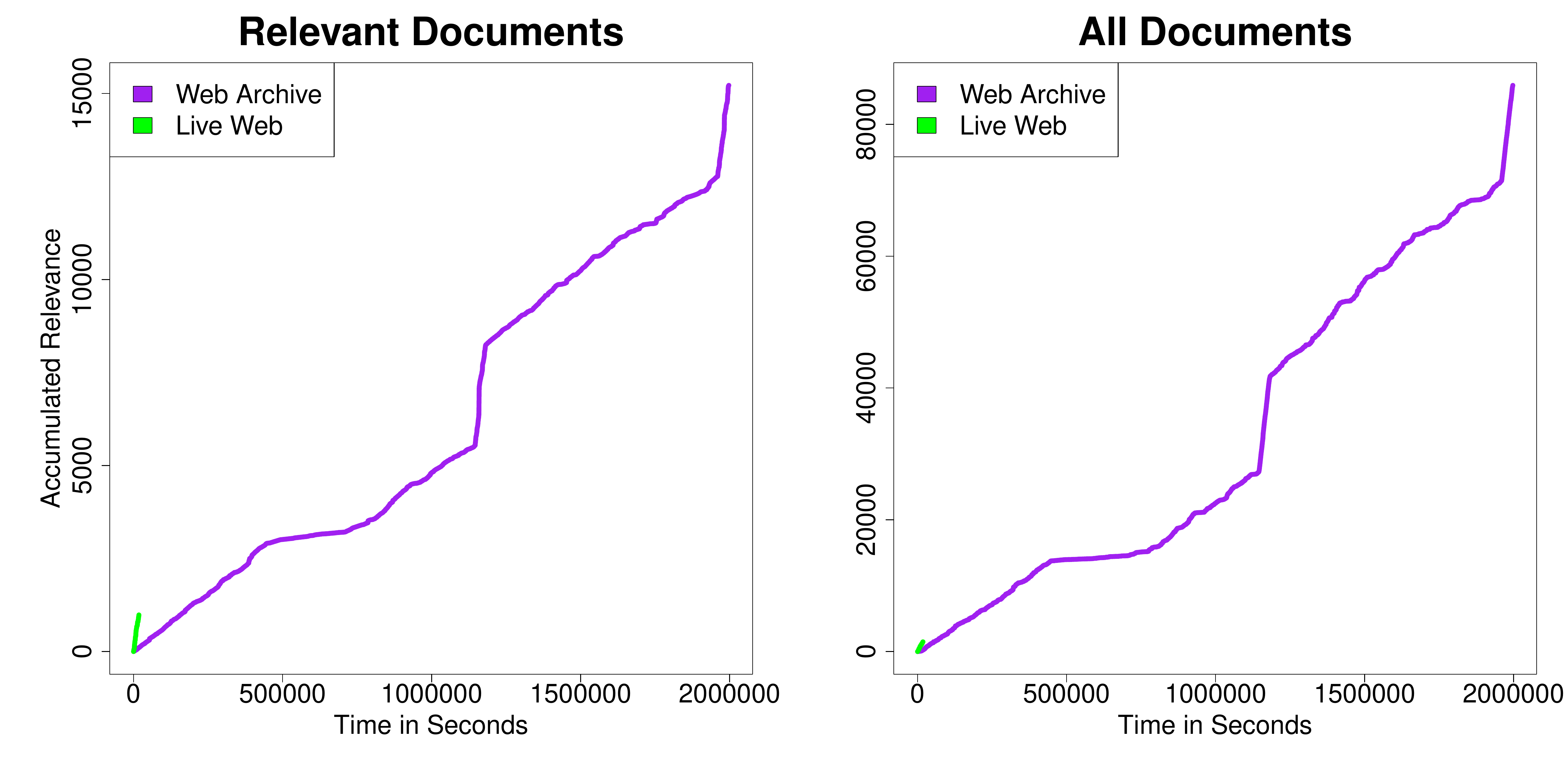}
        \caption{Tucson}
        \label{fig:tu_rel_time}
    \end{subfigure}
    ~
    \begin{subfigure}[t]{0.49\textwidth}
        \centering
        \includegraphics[scale=0.166]{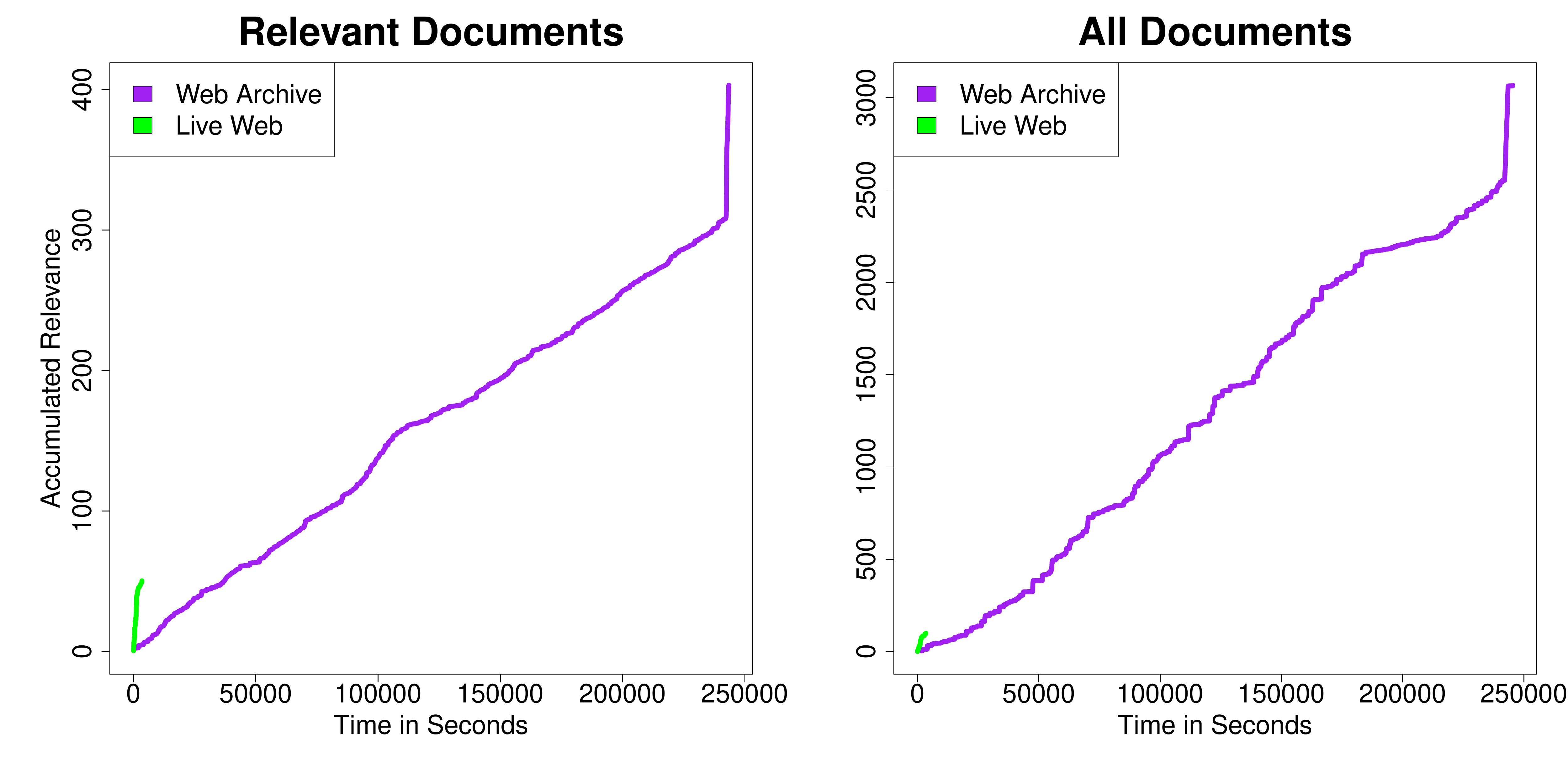}
        \caption{Binghampton}
        \label{fig:bi_rel_time}
    \end{subfigure}
    \caption{Accumulated relevance over time}
    \label{fig:rel_time}
\end{figure*}
\begin{figure*}[h!]
    \centering
    \begin{subfigure}[t]{0.49\textwidth}
        \centering
        \includegraphics[scale=0.166]{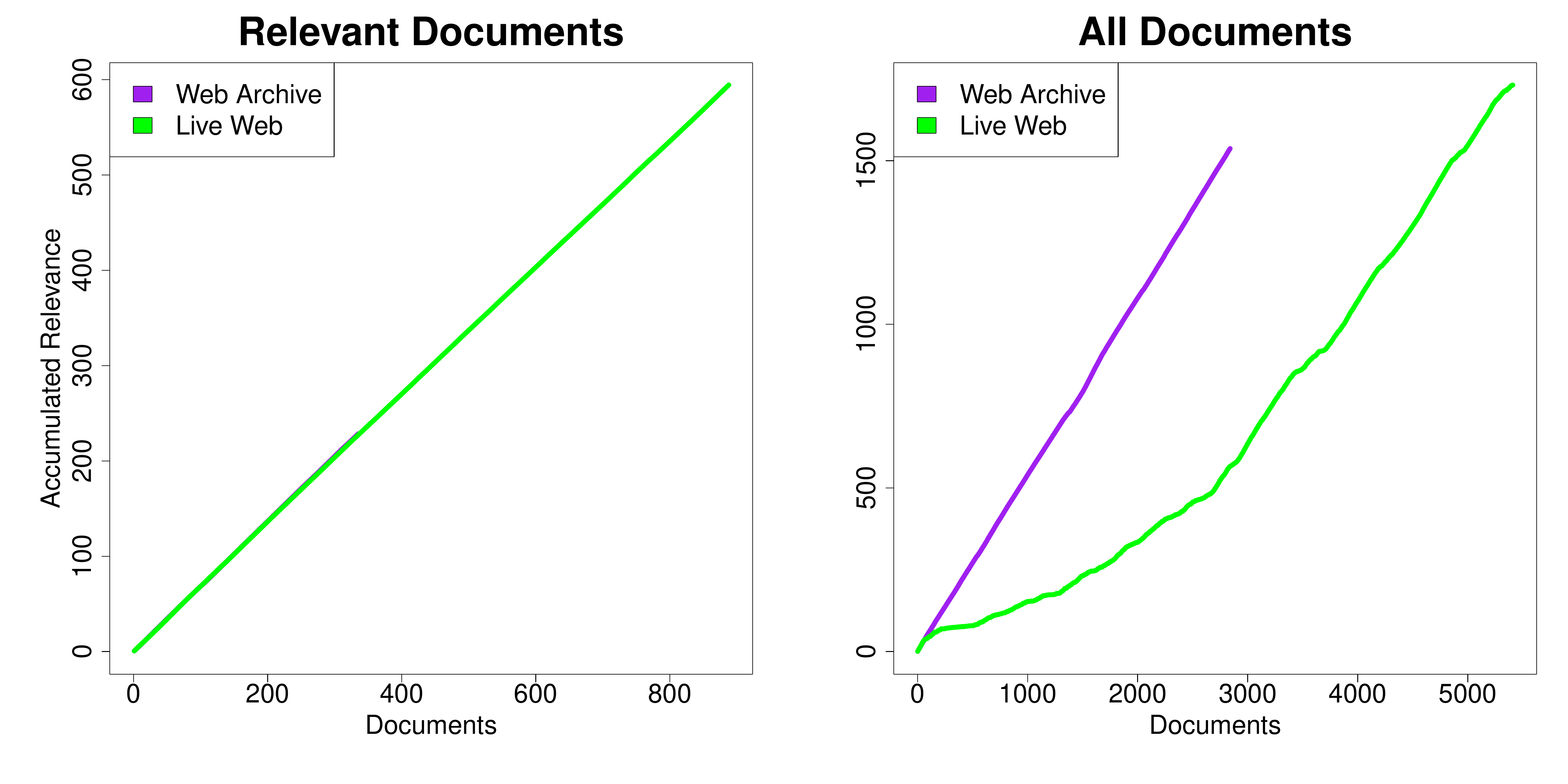}
        \caption{New York City}
        \label{fig:nyc_rel_docs}
    \end{subfigure}
    ~
    \begin{subfigure}[t]{0.49\textwidth}
        \centering
        \includegraphics[scale=0.166]{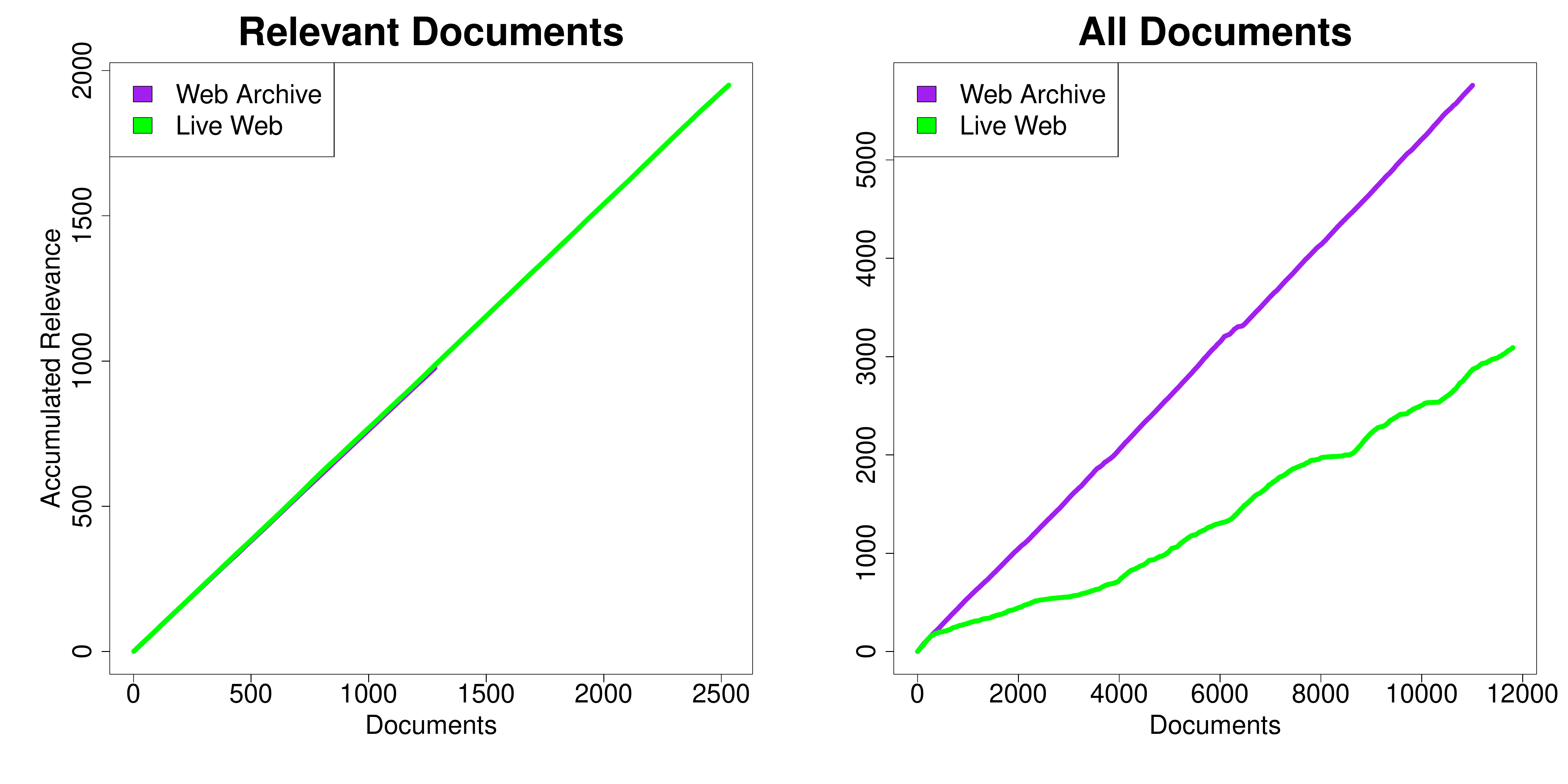}
        \caption{San Bernadino}
        \label{fig:sb_rel_docs}
    \end{subfigure}
    \begin{subfigure}[t]{0.49\textwidth}
        \centering
        \includegraphics[scale=0.166]{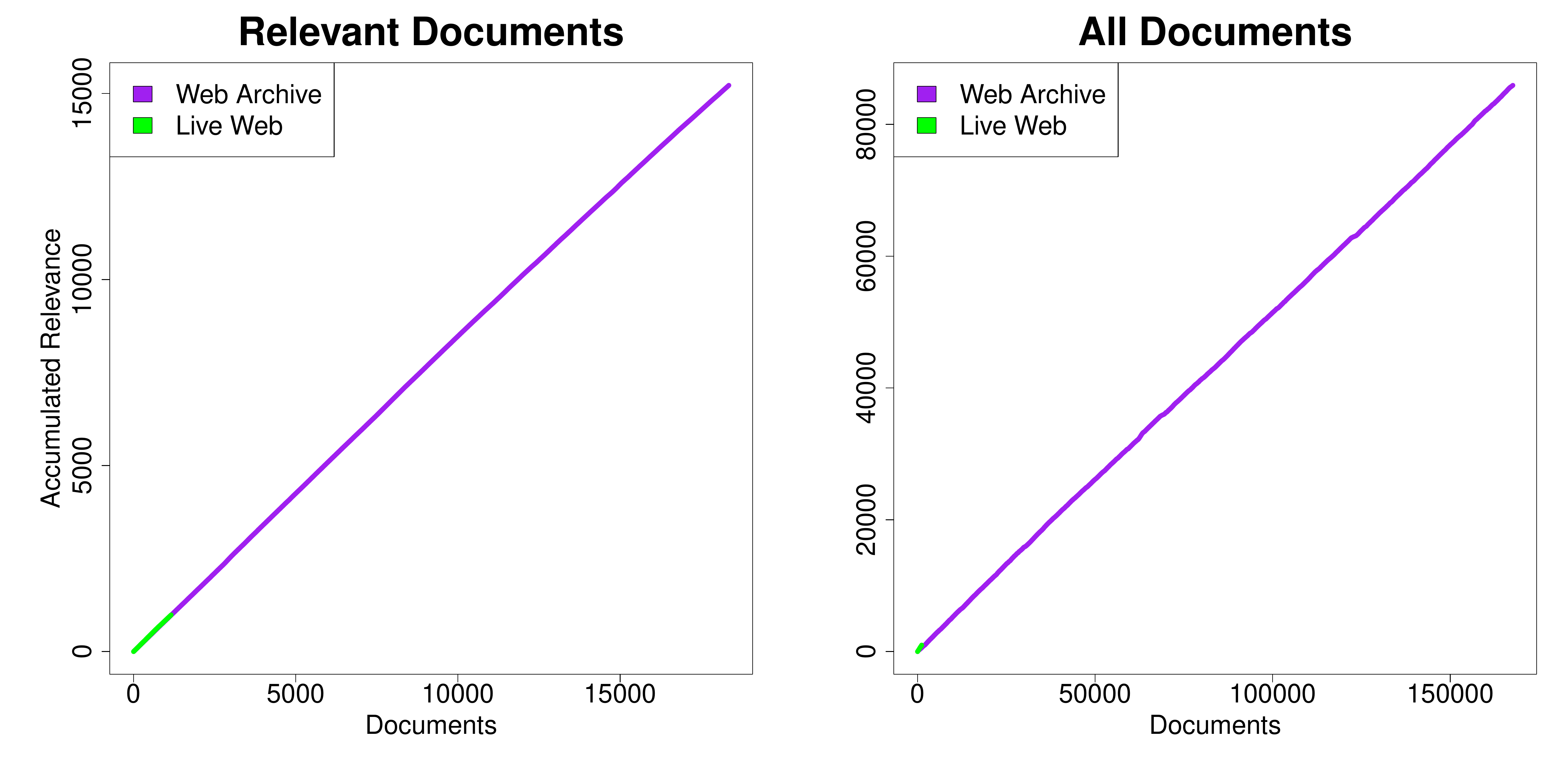}
        \caption{Tucson}
        \label{fig:tu_rel_docs}
    \end{subfigure}
    ~
    \begin{subfigure}[t]{0.49\textwidth}
        \centering
        \includegraphics[scale=0.166]{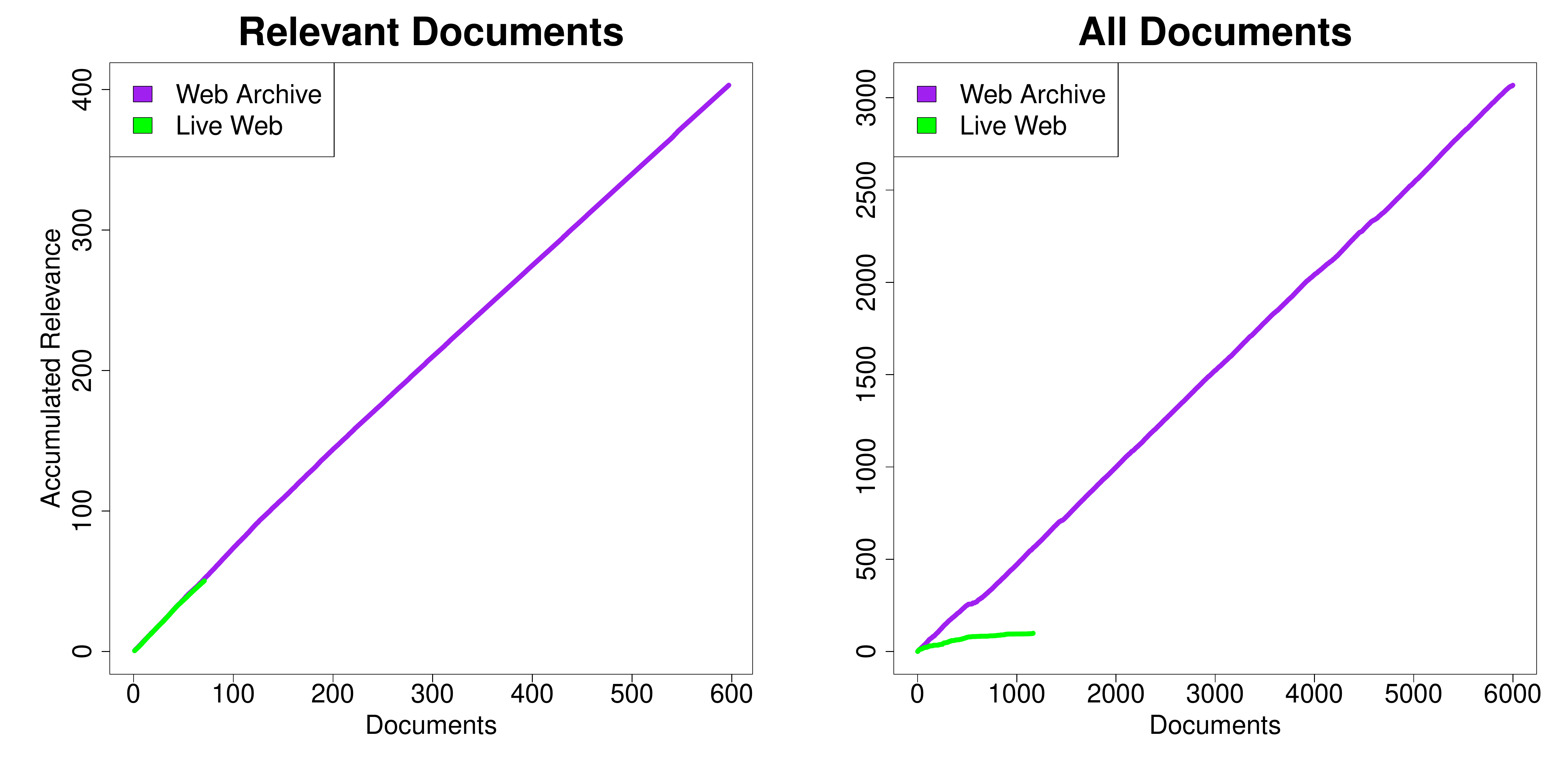}
        \caption{Binghampton}
        \label{fig:bi_rel_docs}
    \end{subfigure}
    \caption{Accumulated relevance over documents}
    \label{fig:rel_docs}
\end{figure*}
\subsection{Accumulated Relevance}
Inspired by the evaluation shown in \cite{gossen:extracting}, we also analyze the accumulated relevance of all crawled resources,
understanding that even resources that do not meet the aggregate relevance threshold still have an aggregate relevance score.
Just like in this related work, we simply add individual $R_{aggr}$ scores of all crawled resources to obtain the accumulated relevance.
Since our crawl stop condition is defined by crawl depth, we are able to show two different analyses of the accumulated relevance.
First, we present the accumulated relevance over elapsed crawl time. We expect the web archive crawl to take longer than the live crawl
as we query the Memento Aggregator for each candidate URI. As described earlier, this results in polling several of the $22$ compliant
web archives, which adds to crawling time.

Figure \ref{fig:rel_time} displays the accumulated relevance (on the y-axis) over time (on the x-axis) for all four events.
The green lines represent the live web crawl and the purple line the web archive crawl. Each subfigure shows two distinct plots. The
plot on the left-hand side shows the data for all resources that were classified as relevant. The plot on the right shows the data
for all crawled resources, including the ones that were crawled because their parent was categorized as relevant but they themselves
had a relevance score below the threshold. These resources, while ``failing'' our threshold test, may still have value for
an event-centric collection and hence they are considered here. Unlike the previous figures, subfigures of Figure \ref{fig:rel_time}
do not distinguish between crawl depths.

Figure \ref{fig:nyc_rel_time} shows the accumulated relevance over time for the NYC crawl. Considering all relevant documents (plot
on the left), we can observe that the accumulated relevance of the live crawl increases very rapidly and that the web archive crawl takes
much longer, as expected, and never reaches the same relevance. The accumulated relevance for all crawled documents (plot on the
right) for the web archive crawl gets closer but still does not reach the accumulated relevance level of the live crawl.
Given the results from the previous section, these observations are not surprising.

Figure \ref{fig:sb_rel_time} shows a similar picture for relevant documents in the SB crawl. However, the data for all crawled
documents is surprising. The web archive crawl takes much longer but eventually surpasses the accumulated relevance level of the live
web crawl. Figures \ref{fig:tu_rel_time} and \ref{fig:bi_rel_time} show an even more dramatic picture.
The accumulated relevance of the live web crawls is quickly surpassed by the web archive crawls.
Given relatively few URIs were obtained in the live crawls (as seen in
Figures \ref{fig:tu_rel_uris} and \ref{fig:bi_rel_uris}), it is not surprising to see these crawls finish rather quickly.
The web archive crawls, again, take significantly longer to complete.
\begin{figure}[t!]
\includegraphics[scale=0.2]{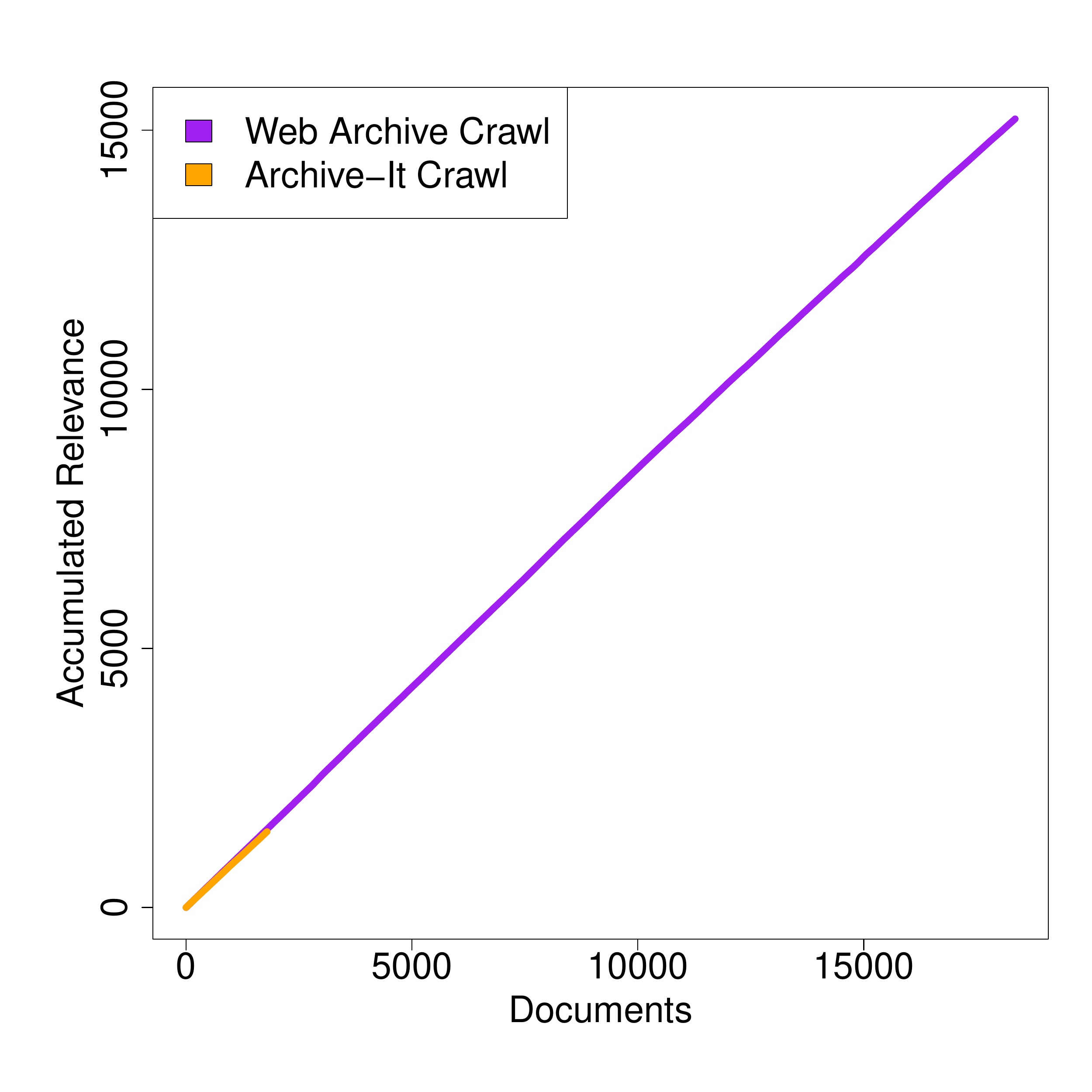}
\caption{TUC web archive crawl vs Archive-It crawl}
\label{fig:tuc_ait}
\end{figure}

The second analysis of the accumulated relevance is over the number of documents crawled. Figure \ref{fig:rel_docs} visualizes
this data in a similar fashion as seen in the previous figures. The data for the NYC crawl is displayed in
Figure \ref{fig:nyc_rel_docs} where we see twice as many relevant documents for the live web than for the web archive crawl. The
accumulated relevance therefore is much higher. When we consider all crawled documents, we also find roughly twice as many resources
in the live crawl and, while the relevance of the web archive crawl is closer, it does not catch up. These data points confirm our
previous findings.

The picture for the live crawl of the SB event (Figure \ref{fig:sb_rel_docs}) is similar to the NYC event. The plot for all 
documents, however, shows an interesting fact: the total number of documents crawled is very similar ($11,806$ for the live web 
vs. $11,007$ for the web archive crawl) while the accumulated relevance of the web archive crawl ends up to be almost twice that 
of the live web crawl.
As previously indicated in Figures \ref{fig:tu_rel_time} and \ref{fig:bi_rel_time}, Figures \ref{fig:tu_rel_docs} 
and \ref{fig:bi_rel_docs} confirm that web archive crawls perform considerably better than live crawls for the TUC and BIN 
events, respectively.
%
%
%
%
%
\begin{figure}[t!]
\includegraphics[scale=0.15]{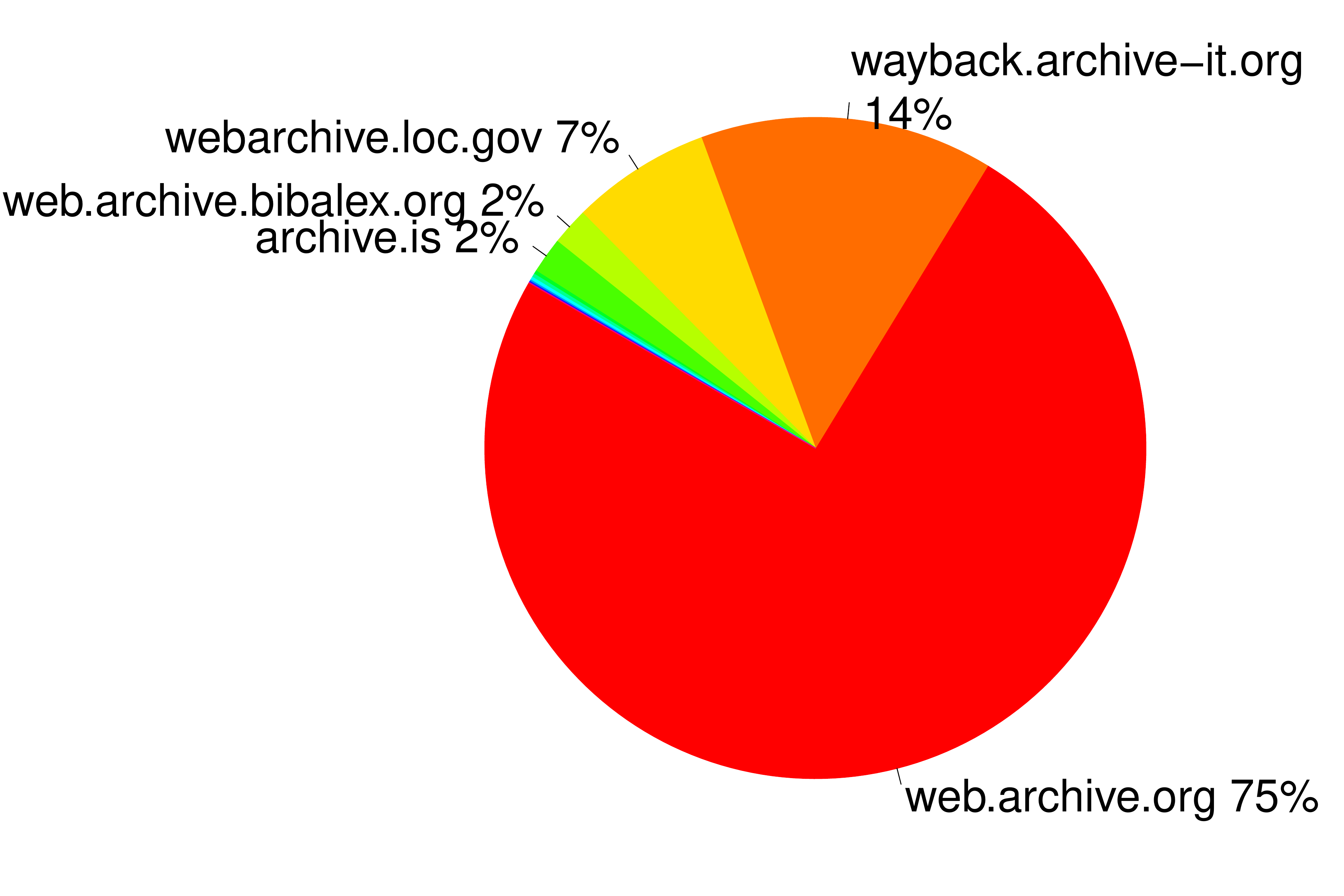}
\caption{Contributions to the TUC web archive crawl}
\label{fig:tuc_archives}
\end{figure}
\section{Comparison to a Manually Created Collection}
We utilized Wikipedia event pages, specifically the URIs of external references as seeds for our crawls. However, a common 
approach for building  event-centric collections from web pages is based on manual suggestion of seed URIs. We are therefore 
motivated to compare our approach with an event collection that was created using manually selected seed URIs. The Archive-It 
service provided by the Internet Archive is frequently used to build such collections. At the time we conducted our experiments, 
the only Archive-It collection that matched one of our events was the Tucson shooting collection, originally created by scholars 
at Virginia Tech. We were able to obtain a copy of the crawled data and compared it to our TUC web archive crawl.

To build this collection, the Archive-It crawler was configured to merely crawl all $1,997$ seed URIs and not go beyond
this crawl depth. In terms of our experiment, this equals to crawl depth $0$ and hence a comparison of relevant URIs per crawl
depth $0..5$ is not applicable. Instead, we compute the accumulated relevance of all crawled resources and compare it to the
data from our web archive crawl. Figure \ref{fig:tuc_ait} shows the results. It is apparent that the Archive-It crawl has 
significantly fewer documents crawled compared to our web archive crawl, an obvious result of the crawl depth constraint. 
However, what is interesting is that the slope of the line is equally steep for both crawls i.e., the orange line (Archive-It 
crawl) and the purple line (our web archive crawl). It would not have been unreasonable to assume that the manually curated 
seed list would result in more relevant URIs crawled than the automatically generated seed list stemming from the references 
of the $DT_{CP}$ version of the Wikipedia page.
%

This comparison raises the question of the level of overlap between the manually curated URIs from the Archive-It
collection and the automatically crawled URIs of our TUC web archive crawl. We classified $1,795$ out of all $1,997$ URIs in 
the Archive-It collection as relevant. On the other hand, we deemed $18,353$ out of $167,641$ crawled URIs in the TUC 
archived crawl relevant. We found that only $92$ URIs overlap in both collections, which indicates that both collections are 
rather disjoint.

Another distinguishing element between these two crawls is the variety of web archives that contribute to the crawl. Given
our framework for crawling the archived web, we are able to crawl archived resources from a total of $22$ web archives. 
Naturally, the Archive-It crawl only stems from one archive. Figure \ref{fig:tuc_archives} shows the distribution of web 
archives that have contributed to our TUC archived crawl. The figure shows the top five contributing archives only, with the 
Internet Archive providing $75\%$ of all Mementos. We note, however, the diversity of other contributing archives. Besides 
resources provided from the Library of Congress and the Library of Alexandria, as shown in Figure \ref{fig:tuc_archives}, 
our crawl further includes resources crawled from the Portuguese, the Icelandic, the UK, and the Northern Ireland Web 
Archives, not labeled in Figure \ref{fig:tuc_archives}.
\section{Conclusion and Future Work}
Inspired by previous work, we were motivated to investigate a focused crawling approach to build event-centric
collections. In this paper we outline our focused crawling framework, detail its methodology, describe its crawling 
process of the live and archived web, and present the results on four unpredictable events.
Our results prove that focused crawling on the archived web is feasible. The Memento protocol and infrastructure play
a vital role in this process. 

Comparing web archive crawls and live web crawls for events, we observe the following patterns:
\begin{enumerate}
\item For rather recent events, such as the NYC event in our experiments, a crawl of the live web results in more total URIs,
more relevant URIs, and a higher level of accumulated relevance over all documents. A web archive crawl is not competitive 
and takes much longer to complete.
\item For events that are less recent but took place in the not too distant past, such as the SAN event in our experiments, 
our results show a mixed pattern. If we consider relevant documents only, the live web crawl outperforms the web archive crawl 
and, as expected, finishes much quicker. However, if time is not a main concern and we can consider all crawled resources, the 
web archive crawl provides more documents that, in aggregate, are more relevant.
\item For events that happened in the more distant past, such as the TUC and BIN events in our experiments, the web archive 
crawl, while taking much longer to complete, returns many more relevant results. A live web crawl does not provide compelling
results. 
\end{enumerate}
The comparison of our web archive crawl on the TUC event with the manually curated Archive-It crawl 
shows that both collections, while distinct in terms of their crawled URIs, are highly relevant to the event.
In addition, we find that the inclusion of an array of web archives clearly provides merit to the collection building.
We therefore suggest that, especially for collections of events that took place in the more distant past, augmenting manually 
curated collections that are based on human-evaluated seed URIs with a focused crawl that is based on the extraction of 
references from Wikipedia pages can be very beneficial.

Our chosen events are constrained in dimensions such as event type, language, location and hence more experimentation is
required to draw general conclusions from our findings. In addition, various aspects of our crawling framework (event vector,
threshold computation, weighting factors) deserve further evaluation in the future.
\section{Acknowledgments}
We would like to express our thanks to Liuqing Li, Edward Fox, and Zhiwu Xie from Virginia Tech for making their
Archive-It collection on the Tucson shooting available for this experiment.
%
%
%
%

%
%
\end{document}